\documentclass[12pt]{article}

\usepackage{graphicx}
\usepackage{graphics}
\usepackage{color}

\def\lsim{\mathrel{\rlap{
\lower4pt\hbox{\hskip-3pt$\sim$}}
    \raise1pt\hbox{$<$}}}     
\def\gsim{\mathrel{\rlap{
\lower4pt\hbox{\hskip-3pt$\sim$}}
    \raise1pt\hbox{$>$}}}     

\begin{document}

\title{ Resonance Structure in the $\gamma\gamma$ Invariant Mass
Spectrum in $p$C- and $d$C-Interactions }

\date{}
\maketitle

\begin{center}
\author{ Kh.U.~Abraamyan$^{a,b,*)}$, M.I.~Baznat$^{c)}$,
A.V.~Friesen$^{a)}$, K.K.~Gudima$^{c)}$, M.A.~Kozhin$^{a)}$,
S.A.~Lebedev$^{d,e)}$, M.A.~Nazarenko$^{a,f)}$,
S.A.~Nikitin$^{a)}$, G.A.~Ososkov$^{d)}$, S.G.~Reznikov$^{a)}$,
A.N.~Sissakian$^{g)}$,
 A.S.~Sorin$^{g)}$, and V.D.~Toneev$^{g)}$\\}

\vspace*{5mm}
 {\it
$a)$ VBLHE JINR, 141980 Dubna, Moscow region, Russia\\
$b)$ Yerevan State University, Yerevan, Armeniya\\
$c)$ Institute of Applied Physics, Kishinev, Moldova\\
$d)$ LIT  JINR, 141980 Dubna, Moscow region, Russia\\
$e)$ Gesellschaft f\"ur Schwerionenforschung, Darmstadt, Germany\\
$f)$ Moscow State Institute of Radioengineering, Electronics and
Automation, 119454 Moscow, Russia\\
$g)$ BLTP  JINR, 141980 Dubna, Moscow region, Russia\\
}
\end{center}

\vspace*{5mm}
\begin{abstract}

Along with $\pi^0$ and $\eta$ mesons, a resonance structure in the
invariant mass spectrum of two photons at $M_{\gamma\gamma}= 360
\pm 7 \pm 9$ MeV is observed in the reaction $d C\to\gamma +
\gamma +X$ at momentum 2.75 GeV/c per nucleon. Estimates of its
width and production cross section are $\Gamma = 63.7 \pm 17.8$
MeV and $\sigma_{\gamma\gamma}=98\pm24^{+93}_{-67}~{\rm \mu b}$,
respectively. The collected statistics amount to $2339 \pm 340$
events of $1.5\cdot 10^6$ triggered interactions of a total number
$\sim 10^{12}$ of $d$C-interactions. This resonance structure is
not observed in $p$C collisions at the beam momentum 5.5 GeV/c.
Possible mechanisms of this ABC-like effect are discussed.\\
 \end{abstract}
$*)$ \textit{abraam@sunhe.jinr.ru}\\
 \textit{Accepted for publication in Phys.Rev.C}

\section{Introduction}

Dynamics of a near-threshold  production of the lightest mesons
and their interactions, especially the pion-pion interaction, are
of lasting interest. A good understanding of the pion-pion
scattering is essential as  it provides a test for the chiral
perturbation theory and the information about quark masses and the
chiral condensate. The two-photon decay of light mesons represents
an important source of information. In particular, the
$\gamma\gamma$ decay of light scalar mesons was considered as a
possible tool to deduce their nature. Also, the scalar-isoscalar
sector is under much debate presently since more states are known
(including possible glueball candidates) than can be fitted into a
single multiplet. Unfortunately, the existing experimental
information from $\pi\pi$ scattering has many conflicting data
sets at intermediate energies and no data at all close to the
 threshold region of interest. For many years this fact has made
it hard to obtain the conclusive results on $\pi\pi$ scattering at
low energies or in the sigma region.

The so-called "ABC effect" is among the oldest and still puzzling
problems. Almost fifty years ago Abashian, Booth, and Crowe
\cite{ABC60} first observed an anomaly in the production of pion
pairs in the reaction $dp \to ^3$He+$2\pi \equiv ^3$He+$X^0$. This
anomaly or ABC effect stands for an unexpected enhancement in the
spectrum of the invariant $\pi\pi$ mass  at masses of about 40 MeV
higher than $2m_\pi$. The subsequent experiments $dp\to ^3$He+$X$
\cite{Bea73}, $pn\to d+2\pi$ \cite{Hea64,Hea69,BNea73}, $dd\to
^4$He+$X$ \cite{Bea76,Wea99}, $np\to d+2\pi$  with neutron beams
\cite{Pea78,Aea79} and even $np\to d+\eta$ \cite{PFW90}
independently confirmed this finding. This anomaly was also
observed in the photoproduction of pion pairs, $\gamma p\to p+X^0$
\cite{R62,DelFea64}. It was revealed that the ABC effect is of
isoscalar nature since a similar effect was not observed in the
$pd\to H^3+X^+$ reaction. The peak positions and widths vary for
different bombarding energies and observation angles. Initially,
the low-mass enhancement was interpreted as being caused by an
unusually strong $s$-wave $\pi\pi$ interaction or as an evidence
for the $\sigma $ meson existence \cite{ABC60} which shortly
before was suggested by Johnson and Teller to provide saturation
and binding in nuclei \cite{JT55}. It is usually accepted now that
this enhancement is not an intrinsic two-pion property since there
is no resonance structure in the $\pi\pi$ scattering amplitude in
this energy range. So any interpretation of the ABC as a real
resonance is very much in doubt (for example, see discussion in
\cite{Pea78}). It is generally believed  that a system like that
has to be associated with two nucleons when two pions (both must
be present) are rescattered off them or both nucleons participate
in elementary $pp\to \pi + X$ reactions (predominantly via
$\Delta$ formation). Actually, the origin of the ABC effect must
be looked for in the formation of light nuclei at intermediate
energies (for a review see Ref. \cite{CP94}).

The presented complicated situation is reflected in the Particle
Data Group (PDG) table \cite{PDT06} where the values quoted for
the sigma mass and width, based on both the pole position and the
Breit-Wigner parameter determinations, are very widely spread. The
estimated mass and half-width are
\begin{equation} m_\sigma -i\Gamma_\sigma
/2=(400-1200) -i(250-500) \ {\rm MeV}. \label{PDT}
\end{equation}
 However, during the last years,  the chiral
perturbation theory and Roy equations led to an accurate
description of $\pi\pi$ scattering at low energies and the precise
determination of the mass and width of the $\sigma$ resonance
\cite{CCL06}:
\begin{equation} m_\sigma=441^{+16}_{-8} \ {\rm
MeV}; \ \ \Gamma_\sigma/2=272^{+9}_{-12.5} \ {\rm MeV}.
\label{CCL}
\end{equation}

All experiments conducted on the ABC issue with the exception of
low-statistics bubble-chamber measurements \cite{BNea73,Aea79}
were inclusive measurements carried out preferentially with a
single-armed magnetic spectrograph for detection of the fused
nuclei. They allow one to find the two-pion invariant mass through
the missing mass. Very recently, exclusive measurements of
reactions $pd\to p+d+\pi^0+\pi^0$ and $pd\to ^3$He$+\pi+\pi$ have
been carried out with complete kinematics in the energy range of
the ABC effect at CELSIUS using the $4\pi$ WASA detector
\cite{Sea06,Cea08}.  The importance of the strongly attractive
$\Delta\Delta$ channel was noted. Surprisingly, the basic $pp\to
pp\pi^0\pi^0$ reaction in the $\Delta\Delta$ region also shows an
ABC-like low-$\pi\pi$ mass enhancement which deserves special
attention. It confirms the earlier result in \cite{Yea01} where
the analyzing powers and cross sections for the ABC enhancement
production  were measured for the reaction $\vec{p}p\to p+p+X^0$
in the missing mass range $m_\pi^0 < M_{X^0}<m_\eta$. However, the
big difference was observed in the width of the resonance cross
section and it was concluded that the observed width in the
isoscalar channel is not obviously just a simple result of the
binding between the two $\Delta$ states. It rather signals more
complicated configurations in the wave function of the
intermediate state, as would be expected for a nontrivial dibaryon
state \cite{Cea08}.

This work aims to study whether this low-$\pi\pi$ mass anomaly can
survive in heavier systems in the $\gamma\gamma$ channel. The
paper is organized as follows. After a brief description of the
experiment and experimental setup, the structure of measured
invariant mass spectra  of photon pairs is analyzed in Sect.2. As
a cross-check, a similar analysis is carried out in Sect.3 but
within the wavelet method. To elucidate the nature of the peak at
$M_{\gamma\gamma}\approx (2-3)m_\pi$, different mechanisms of the
observed $\gamma\gamma$ pair enhancement are discussed in Sect.4.
In Sect.5, experimental estimates for production cross sections
and widths of $\eta$ mesons and hypothetical $R$ resonance are
given. The main inferences of the paper are presented in the
Concluding Section.

\section{Experiment}
\subsection{General layout}
The data acquisition of production of neutral mesons and
$\gamma$-quanta in $p$C and $d$C interactions  has been carried
out with internal beams of the JINR Nuclotron \cite{2,3}. The
experiments were conducted with internal proton beams at momentum
5.5 GeV/c incident on a carbon target and with $^2$H, $^4$He beams
and internal C-, Al-, Cu-, W-, Au-targets at moments from 1.7 to
3.8 GeV/c per nucleon. For the first analysis the data with the
maximal statistics, $p$C- and $d$C-interactions, were selected.
The first preliminary results on $d{\rm C} (2~AGeV)$ collisions
were reported in \cite{3} where some indication on unusual
structure in the photon-photon invariant spectrum has been
obtained.

\begin{figure}[h]
\centerline{ \includegraphics[width=8.6cm]{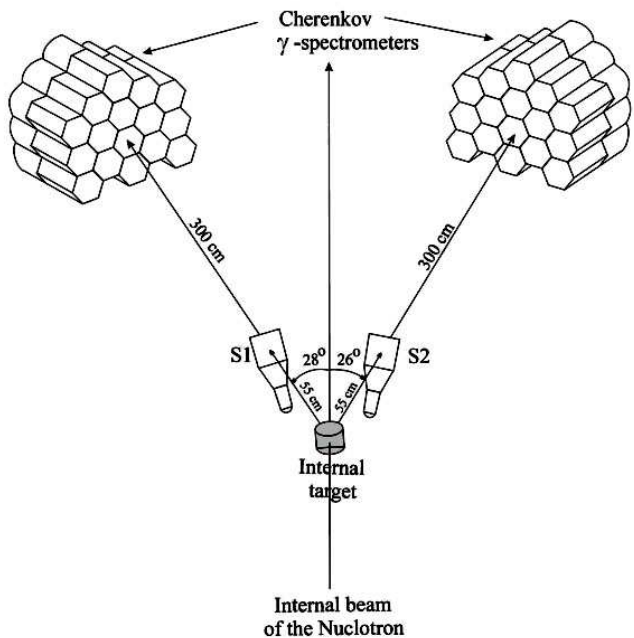}}
 \caption{The schematic drawing of the experimental PHOTON-2 setup.
 The $S_1$ and $S_2$ are scintillation counters.  }
  \label{set}
\end{figure}

The presented data concern reactions induced by deuterons with a
momentum 2.75 GeV/c per nucleon and by protons with 5.5 GeV/c.
Typical deuteron and proton fluxes were about $ 10^9$ and $2\cdot
10^8$ per pulse, respectively. The electromagnetic lead glass
calorimeter PHOTON-2 was used to measure both the energies and
emission angles of photons. The results obtained in earlier
experiments with this setup are published in \cite{Ab94}. The
experimental instrumentation is schematically represented in
Fig.\ref{set}.

 The PHOTON-2 setup includes 32 $\gamma$-spectrometers of lead
glass and scintillation counters $S_1$ and $S_2$ of $2\times 15
\times 15~cm^3$ used in front of the lead glass for the charged
particle detection \cite{Ab94,Ab89,Kha85}.

\begin{table}[h]
\caption{The basic parameters of the lead glass hodoscope}
\label{tab1} \vspace*{3mm}\hspace*{15mm}
\begin{tabular}{|l|l|}
\hline
Number of lead glasses &32 TF-1, total wight 1422 kg\\
 Module cross section  & r=9 cm of insert circumference\\
 Module length          & 35 cm, 14 R.L. \\
 Spatial resolution     & 3.2 cm  \\
 Angular resolution     & 0.6$^0$  \\
 Energy resolution      & $(3.9 /\sqrt E+0.4)\%$, E~[GeV] \\
 Gain stability         & (1-2)$\%$  \\
 Dynamic range          &  50 MeV - 6 GeV \\
 Minimum ionizing signal  & 382 $\pm$4 MeV of the photon\\
                          & equivalent \\
 Total area             & 0.848  $m^2$ \\
 \hline
\end{tabular}
\end{table}

 The center of the front surfaces of the lead glass
hodoscopes is located at 300 cm from the target and at angles
25.6$^\circ$ and 28.5$^\circ$ with respect to the beam direction.
This gives a solid angle of 0.094 {\it sr} (0.047 {\it sr} for
each arm). Some details of the construction and performance of the
lead glass hodoscope are given in Tabl.\ref{tab1}.  The internal
target consists of carbon wires with the diameter of 8 micron
mounted in a rotatable frame. The overall construction is located
in the vacuum shell of the accelerator.

 Before the experiment the energy calibration of the lead glass
counters has been carried out  with 1.5 GeV/c per nucleon
deuteron-beam at the JINR synchrophasotron \cite{Ab89}. The
long-term gain stability was continuously monitored for each of
the lead glass modules with 32 NaI(Tl) crystals doped with
$^{241}$Am sources.

The modules of the $\gamma$-spectrometer are assembled into two
arms of 16 units. These modules in each arm are divided into two
groups of 8 units. The output signals of each group from 8
counters are summed up linearly and  sent to the inputs of four
discriminators ($D_i$). In this experiment all  the discriminator
thresholds were at the level of 0.4 GeV. Triggering takes place
when there is a coincidence of signals from two or more groups
from different arms: $ (D_1+D_2)\times(D_3+D_4)$. In realizing the
trigger conditions the amplitudes of all 32 modules were recorded
on a disc. The dead time of data acquisition is about 150 $\mu s$
per trigger. The mean rate of triggering was about 330 and 800
events per spill in $d$C and $p$C reactions, respectively.
Duration of the irradiation cycle is 1 second. Totally about
$1.52\times10^6$ and $1.06\cdot 10^6$ triggers were recorded
during these experiments.

\subsection{Event selection}\label{sec2.2}

Photons in the detector are recognized as isolated and confined
clusters (an area of adjacent modules with a signal above the
threshold) in the electromagnetic calorimeter. The photon energy
\begin{figure}[h]
\centerline{\includegraphics[width=8cm]{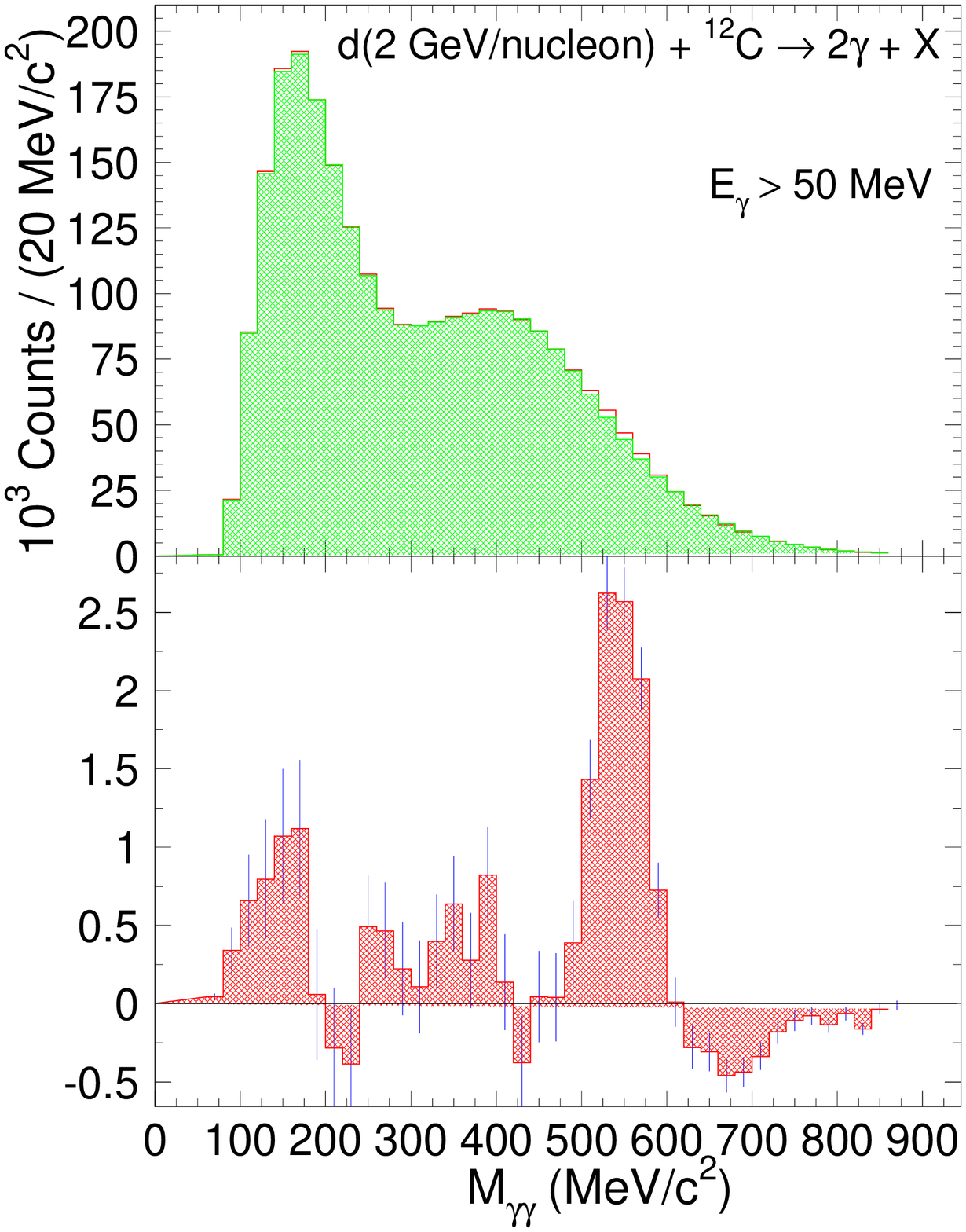}
\includegraphics[width=8cm]{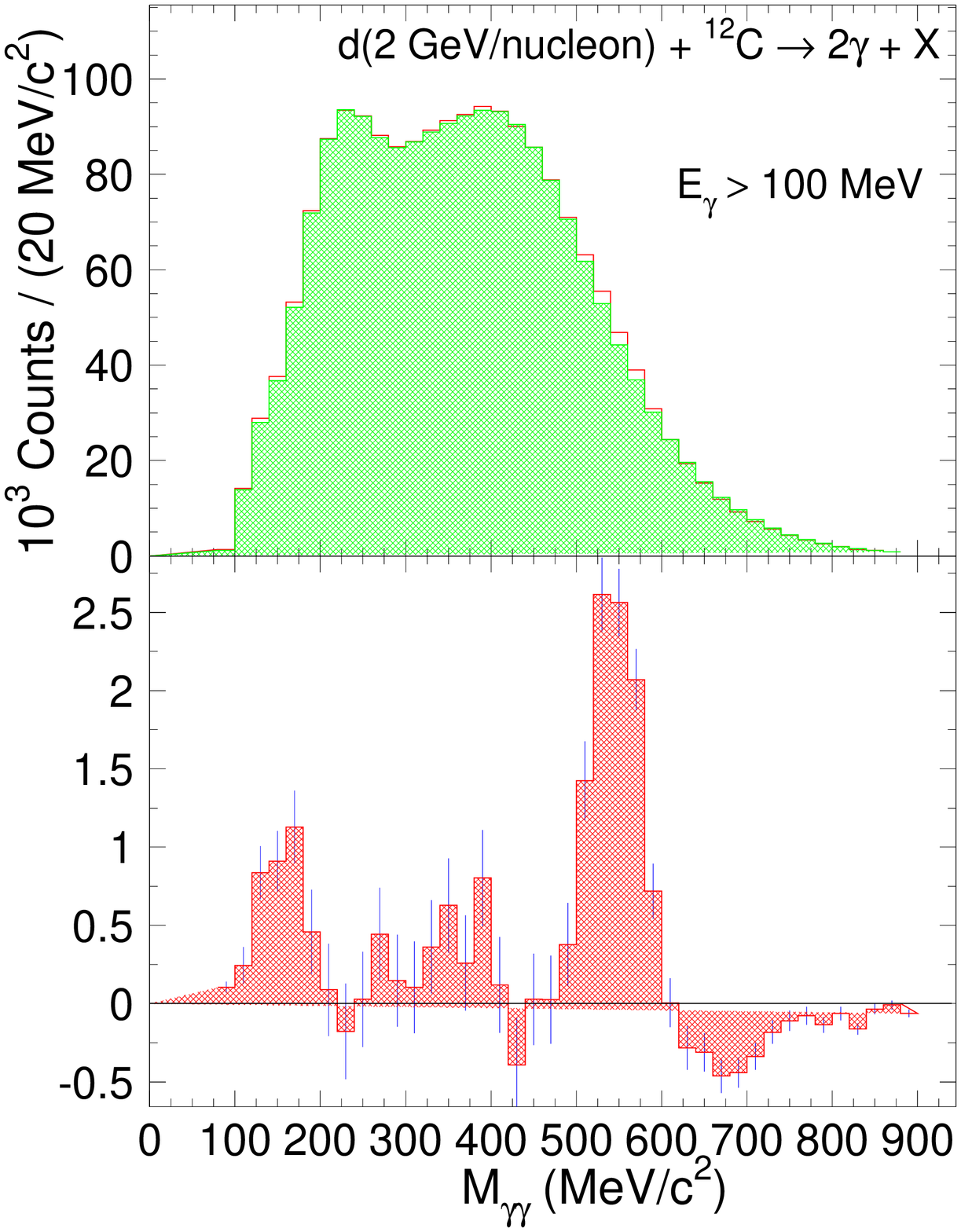}}
\caption{(color online) Invariant mass distribution of
$\gamma\gamma$ pairs from the reaction $d{\rm C}\to \gamma +\gamma
+X$ at 2.75 GeV/c per nucleon for two values of the cut energy of
photons. The top shaded histograms show the background
contribution. The bottom histograms are invariant spectra after
the background subtraction. The auxiliary normalization factor
$K_{norm}$ (see below) is 0.9947 for left figure (the cut energy
is 50 MeV) and $K_{norm} = 0.993$ for right figure (the cut energy
is 100 MeV).}
  \label{f3}
\end{figure}
is calculated from the energy of the cluster. Cluster
characteristics were tested by comparison with Monte-Carlo
simulations of electron-photon showers in Cherenkov counters by
means of the program package EMCASR~\cite{EMCASR}. The results
obtained earlier with extracted ion beams of the JINR
Synchrophasotron have demonstrated a good agreement between
experimental and simulated data~\cite{Ab94}. Assuming that the
photon originates from the target, its direction is determined
from the geometrical positions of constituent modules weighted
with the corresponding energy deposit in activated modules.
\begin{figure}[h]
\centerline{\includegraphics[width=10cm]{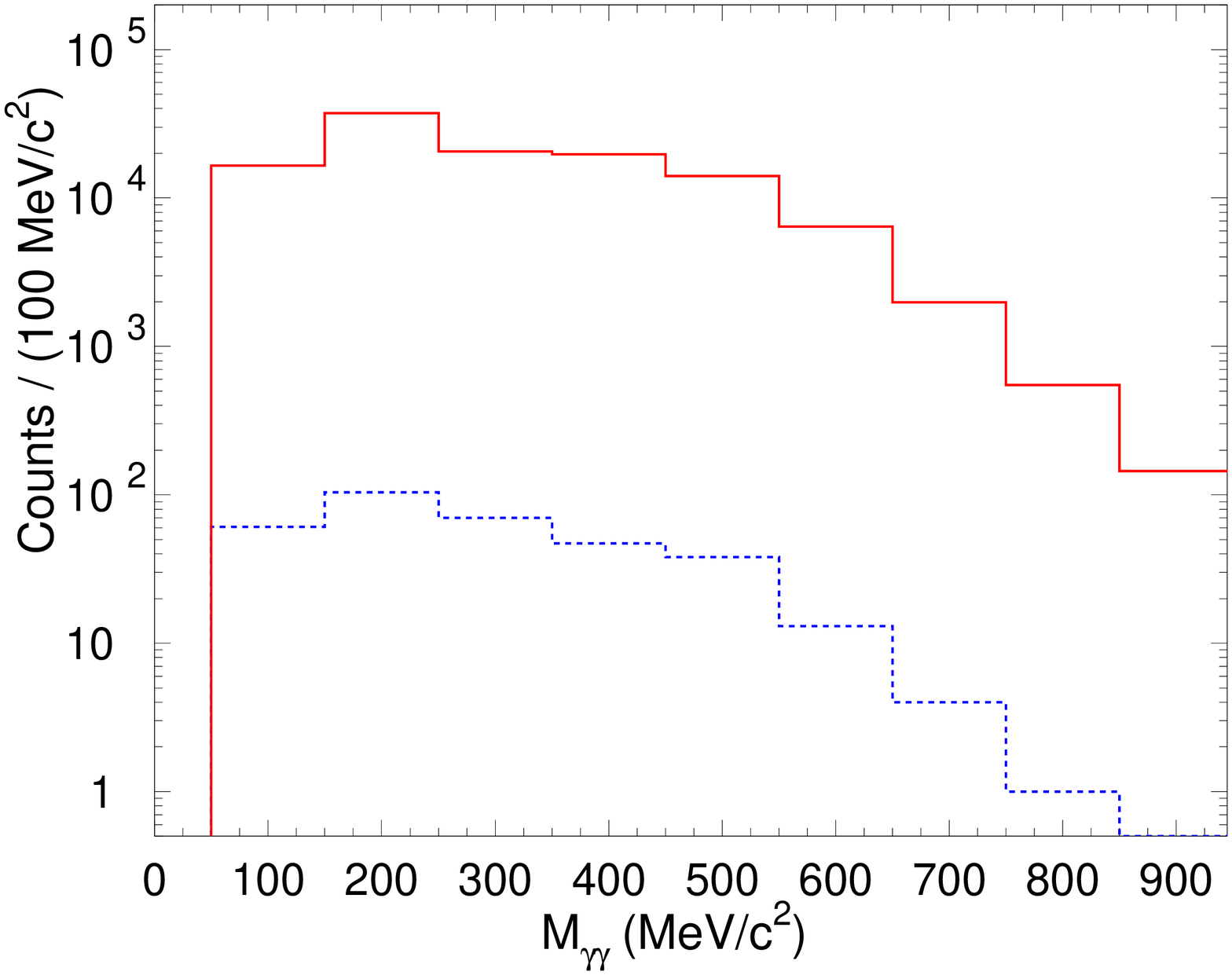}}
\caption{(color online) Invariant mass distributions of
$\gamma\gamma$ pairs in two different runs of measurement under
condition $E_\gamma \geq 50$ MeV: with the empty target (dashed
histogram) and with the internal carbon target (solid histogram)
in the reaction $d{\rm C}=\gamma + \gamma + X$ at 2.75 GeV/c per
nucleon. }
  \label{empty}
\end{figure}

The invariant mass distributions of photon pairs  (from different
arms of the spectrometer) are shown in Fig.\ref{f3}. The dominant
part of distributions (two upper panels) comes from the $\pi^0\to
\gamma\gamma$ decay. These photons in combination with others
result in a huge background  which mask the expected $\eta \to
\gamma\gamma$ decay. Other sources of background are charged
particles as well as neutrons and particles from a general
background in the accelerator hall. Contributions of the given
sources were estimated by special measurements with and without
veto-detectors $S1$ and $S2$  and by comparison of data obtained
at different beam intensities. The contribution to the total
combinatorial spectrum  of charged particles, neutrons and
particles from a general background in the accelerator hall is
less than $10\%$ and becomes negligible ($<1\%$) after subtraction
of mixing event background (see below). As follows from
Fig.\ref{f3}, the high-energy cut of photons, $E_\gamma>100$ MeV,
allows one to improve the signal-to-background ratio.  The
contribution of the general background in the experimental hall
was estimated from the measurements with the empty target, see
Fig.\ref{empty}. Two runs carried out for 125 accelerator cycles
of the deuteron beam (about $~10^{11}$ deuterons in every run)
result in $N_{\gamma\gamma} =$ 117428 and 338 photon pairs for the
case with and without target, respectively. So this source
contributes less than $1\%$ and is quite smoothly distributed with
respect to $M_{\gamma\gamma}$.

\begin{figure}
\vspace*{-8mm}
\includegraphics[width=8cm]{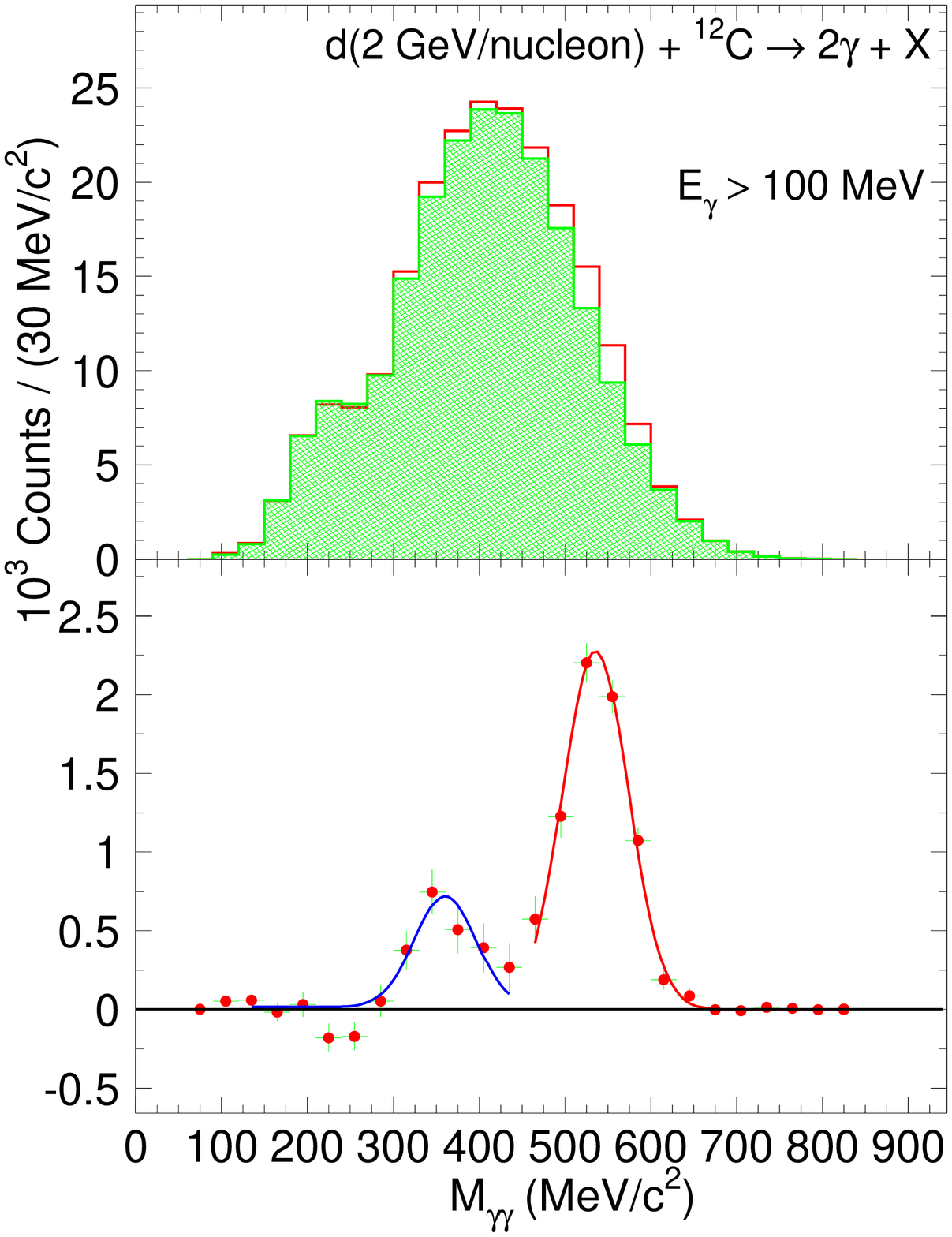}
 \includegraphics[width=8cm]{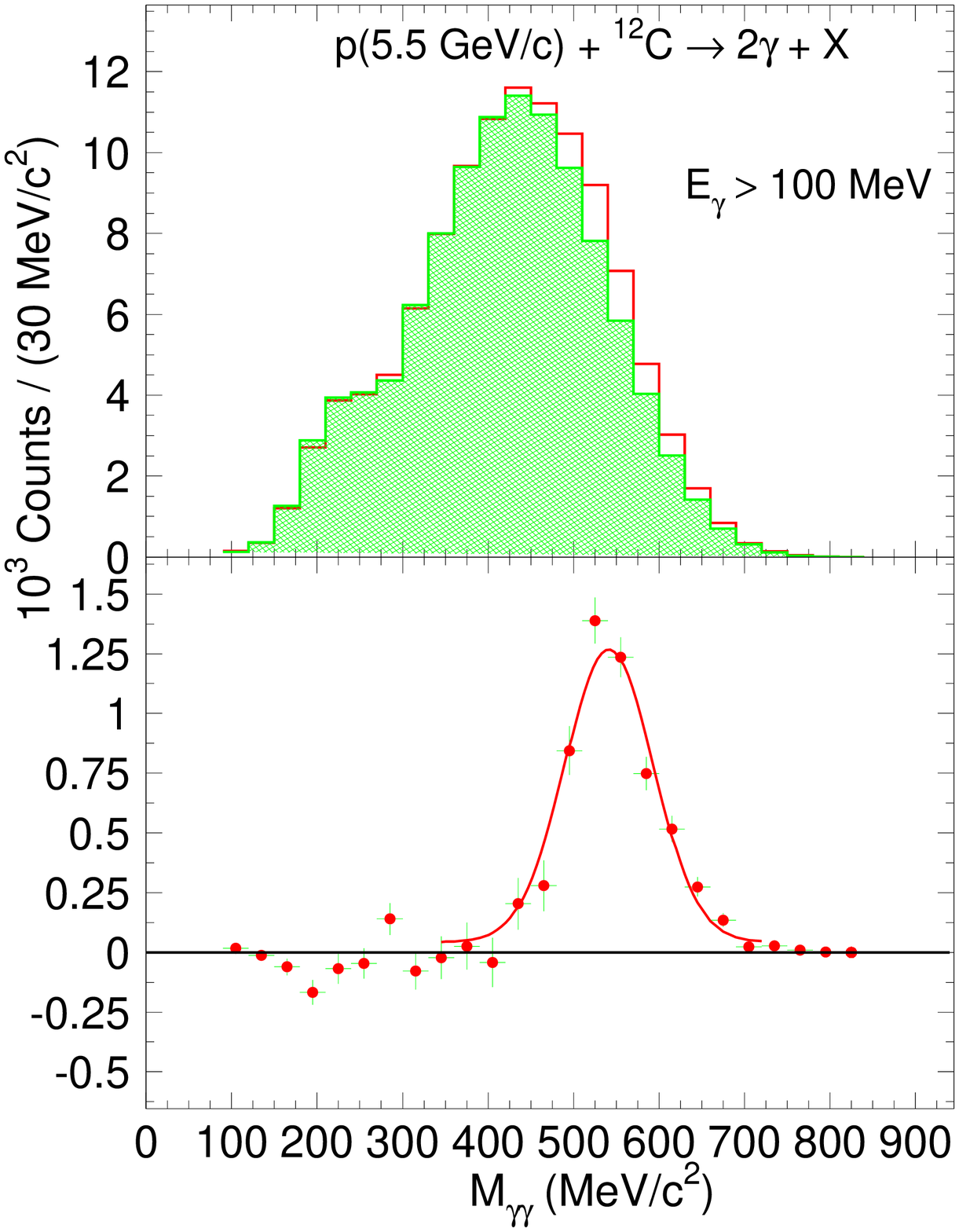}
  \caption{(color online) Invariant mass distributions of $\gamma\gamma$
  pairs satisfying criteria $(1)-(3)$   without (upper panels) and with
(bottom panels) the background subtraction
 obtained for the reaction $d{\rm C}\to \gamma +\gamma
+X$ at 2.75 GeV/c per nucleon (left) and $p$C collision at 5.5
GeV/c (right), respectively. The curves are the Gaussian
approximation of experimental points (see the text). The values of
$K_{norm}$ are 0.958 for $dC$ and 0.952 for $pC$.
  }
  \label{trig}
\end{figure}

To see a possible structure of the invariant mass spectra, a
background should be subtracted. The so-called event mixing method
was used to estimate the combinatorial background: a photon in one
event from the first arm  is combined with a photon in other
events from the second  arm. This background
 was subtracted from the invariant mass distributions (see
bottom panels in Fig.\ref{f3}). The background normalization was
carried out in two steps. First, the background is normalized to
the total pair number (see [21]). Naturally that at the event
mixing the resonance maxima are not reproduced disturbing the
overall normalization. At the second step this shortcoming of the
background estimate is corrected by an auxiliary factor $K_{norm}$
(which differs by few $\%$ from 1) obtained by iterating treatment
of the resonance contribution to the spectrum. The negative values
in the high-mass range arising at the subtraction of the
mixing-event background are caused by the energy conservation law
which may be violated when $\gamma$-quanta are taken from
different mixing events. To decrease its influence the energy sum
of $\gamma$-quanta in both individual events and mixing
$\gamma\gamma$ pairs is restricted (see below criterion (3)). A
clear peak from the $\eta$ decay and a remnant of the suppressed
$\pi^0$ resonance are clearly observable. Note that between them
there is some additional structure which will be clarified below.

Systematic errors may be due to uncertainty in measurements of
$\gamma$ energies and inaccuracy in estimates of the combinatorial
background. The method of energy reconstruction of events is
described in detail in Refs. \cite{Ab94,Ab89}. Possible
overlapping effect was studied at higher intensities in CC
collisions at the 3.7 GeV per nucleon beam energy. It may result
in about $6\%$ displacement of the total mass spectrum. General
influence of the overlapping effect in this experiment is very
small. One of the criteria of accuracy of energy reconstruction is
the conformity of the peak positions  corresponding to the known
particle mass values. As is seen in Fig.\ref{f3}, the position of
peaks corresponding to $\eta$- and $\pi^0$-mesons  is in
reasonable agreement with the table values of their masses. A more
precise determination of the position of peaks requires
minimization of systematic errors in describing the background
which arise, in particular, due to the violation of the
energy-momentum conservation laws in selecting $\gamma$-quanta by
random sampling from different events (see also below).

The selection criteria can be made harder by imposing additional
trigger conditions. For a background suppression  and minimization
of systematical errors due to violation of conservation laws the
following selection criteria were used:

 (1) the number of photons in an event, $N_\gamma =2$;

 (2) the energies of photons, $E_\gamma \ge 100$ MeV;

 (3) the summed energy in real and random events $\le 1.5$ GeV.
The criterion (1) suppresses the combinatorial background and
minimizes systematical errors arisen due to the violation of
photon topology  at event mixing because events with a different
number of photons $N_\gamma$ have different  angular and energy
distributions. As noted above, the criterion (2) improves the
signal-to-background ratio. The criterion (3) also allows one to
minimize systematical errors at event mixing at an insignificant
($\sim 3\%$) loss of events. The result of this triggering is
shown in Fig.\ref{trig}. Under the selected condition the $\pi^0$
peak is practically absent. Therefore, $\pi^0$-mesons were mainly
detected in events with $N_{\gamma}
> 2$ (a minimal opening angle of the $\gamma$ pair detected by the
setup equals $42^\circ$). In contrast, the $\eta$ is seen very
distinctly  with the width defined by the experimental resolution
in the mass. In addition, in this reaction $d{\rm C}\to\gamma
+\gamma +X$  a pronounced peak is observed in the interval 300-420
MeV of the invariant mass of two-photon spectrum which will be
named below the R-resonance.

\begin{table}
\caption{Fit parameters of the Gaussian distribution} \label{tab2}
\vspace*{3mm}\hspace*{5mm}
\begin{tabular}{|c|c|c|}
\hline
       & $d{\rm C}$  & $d{\rm C}$ \\
       &$ 165\leq M_{\gamma\gamma}\leq 435$ MeV & $465\leq M_{\gamma\gamma}\leq 825$ MeV \\
\hline
 $y_0$ & -1.94 $\pm$ 1.31       & -0.004 $\pm$ 0.046   \\
 $N_0$ &  2623 $\pm$ 472      & 7329 $\pm$ 295      \\
 $w_{\rm measur}$, MeV   & 41.3 $\pm$ 7.2  & 38.5 $\pm$ 1.7 \\
 $M_0$,  MeV/$c^{2}$ & 362.0 $\pm$ 6.9     & 535.7 $\pm$ 1.9     \\
$\chi^2~/$degrees of freedom &  9.06/6     & 7.38/9   \\
\hline
       & $p{\rm C}$ & $p{\rm C}$ \\
       & $ 345\leq M_{\gamma\gamma}\leq 645 $ MeV & $405\leq M_{\gamma\gamma}\leq 645$ MeV \\
\hline
 $y_0$ &  1.44 $\pm$ 1.86 & 1.30 $\pm$ 3.07 \\
 $N_0$ & 5283 $\pm$ 560 & 4804 $\pm$ 673 \\
 $w_{\rm measur}$, MeV   & 51.6 $\pm$ 4.1 & 41.9 $\pm$ 4.2\\
 $M_0$,  MeV/$c^{2}$  & 541.5 $\pm$ 2.5 & 536.6 $\pm$ 2.6 \\
$\chi^2~/$degrees of freedom &    16.10/7   &  3.13/4         \\
\hline
\end{tabular}
\end{table}

However, under similar conditions only $\eta$ is seen in the $p$C
collisions. The observed peaks were approximated independently by
the Gaussian:
\begin{eqnarray}
\frac {dN} {dM_{\gamma\gamma}}=y_0+\frac {N_0} {w_{\rm
measur}\sqrt{2\pi}} \exp\left({-\frac {(M_{\gamma\gamma}-M_0)^2}
{2w_{\rm measur}^2}}\right).
 \label{eq2}
 \end{eqnarray}

The additional shift-parameter $y_0$ is introduced in
eq.(\ref{eq2}). The values of the obtained fitting parameters are
given in Tabl.\ref{tab2}.

 The signal-to-background ratios for the invariant mass
intervals of 300-420 MeV and 480-600 MeV (the vicinity of the
$\eta$-meson mass) are $2.5\cdot 10^{-2}$ and $1.4\cdot 10^{-1}$,
respectively. For comparison, analogous values without the
background suppression (without the selection criteria (1)-(3))
are $(4.0\pm 1.4)\cdot 10^{-3}$ and $3.2\cdot 10^{-2}$.
\begin{figure}
\includegraphics[width=8cm]{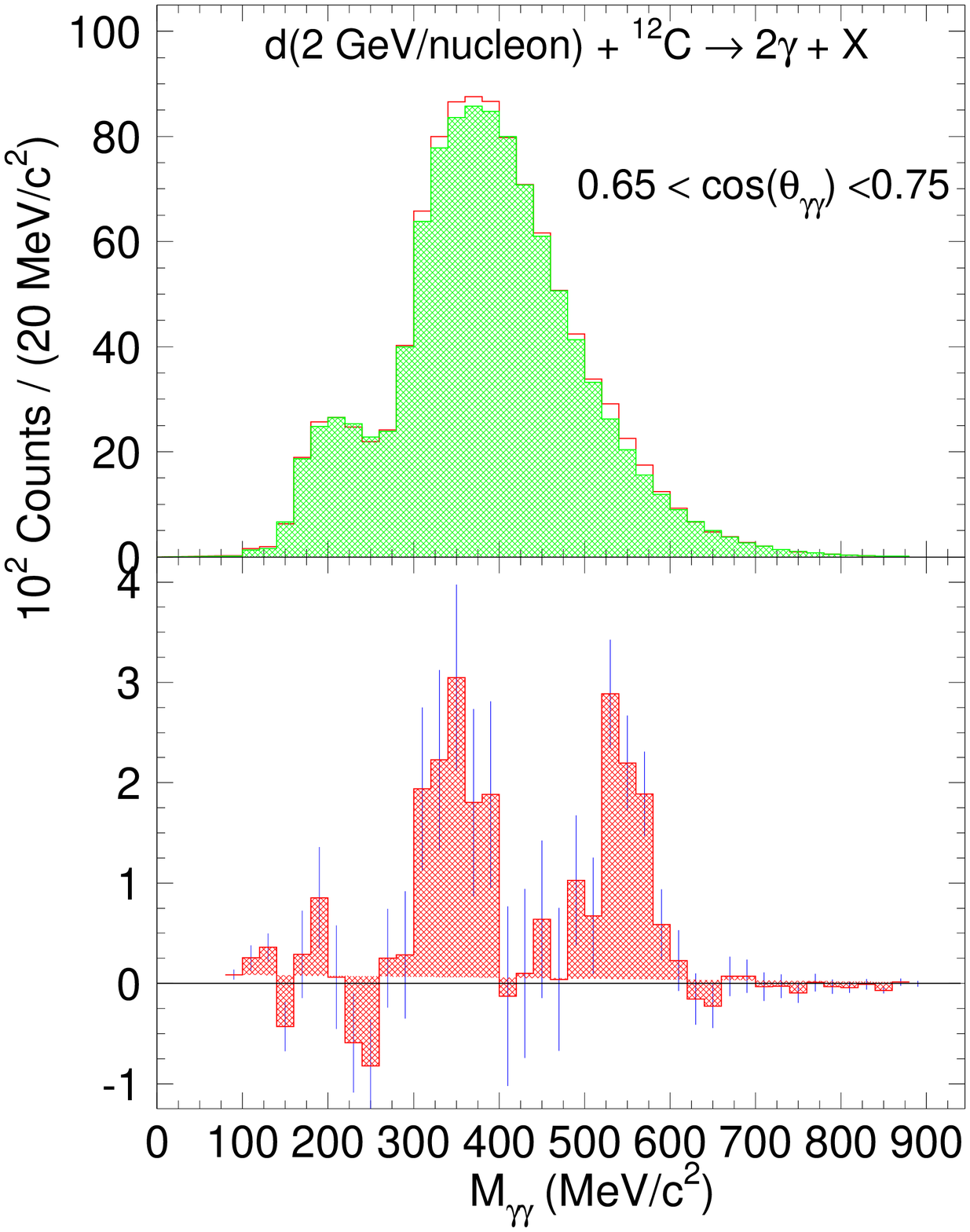}
\includegraphics[width=8cm]{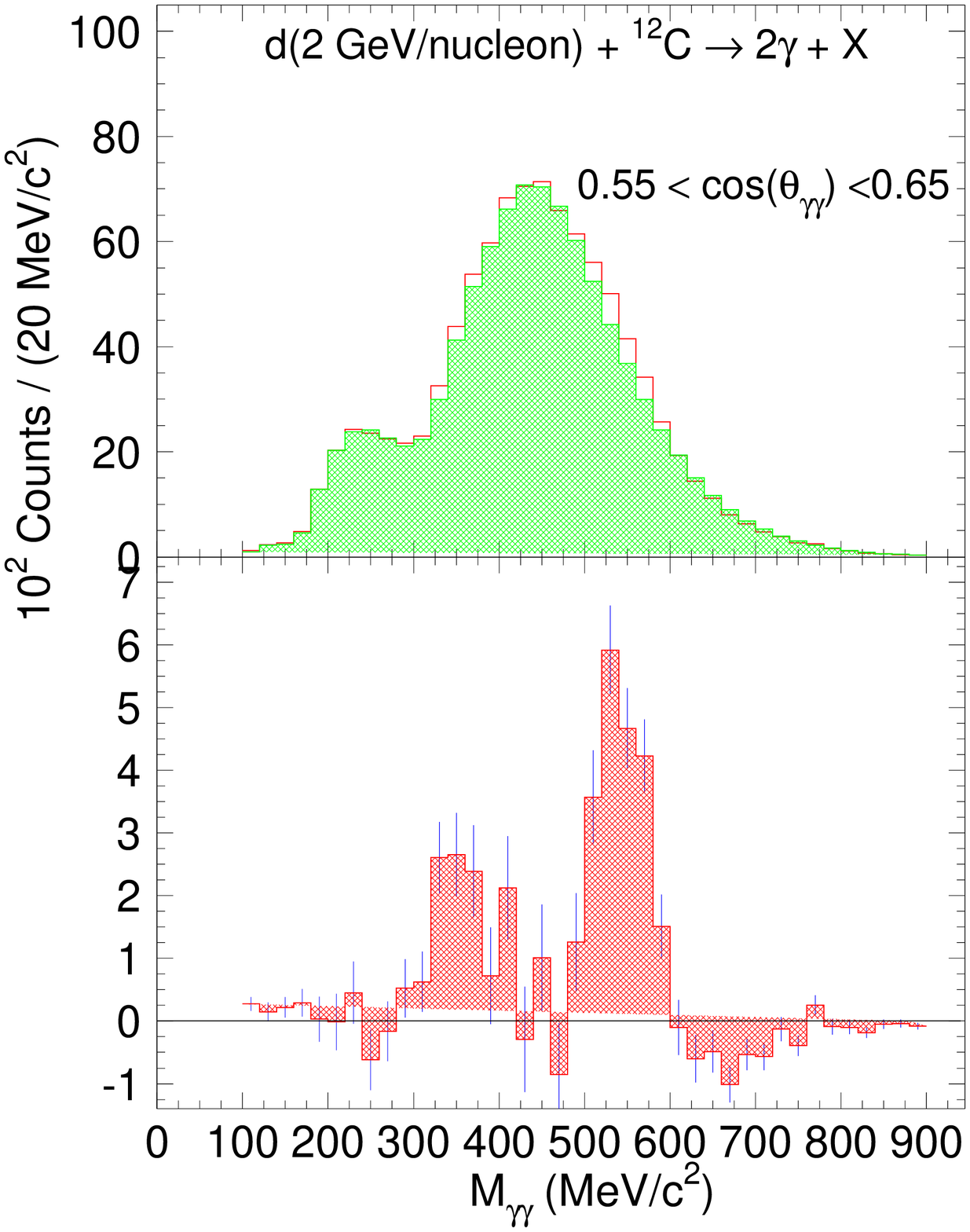}
  \caption{(color online)
The invariant mass distributions of two photons for the opening
angles $0.55<\cos \Theta_{\gamma\gamma}<0.65$ (left) and
$0.65<\cos \Theta_{\gamma\gamma}<0.75$ (right) under the selection
criteria $(1)\div (2)$. The values of $K_{norm}$ are 0.98 (left)
and 0.97 (right).
  }
  \label{op_ang}
\end{figure}

Thus, as follows from Tabl.\ref{tab2}, the position and width of
the peak corresponding to $\eta$-meson are in good agreement with
values from the PDG table (systematic errors do not exceed 2$\%$)
and the spectrometer  mass resolution. The total number of
detected events in the $d$C reaction for the $\eta$-meson region
450-660 MeV after background subtraction is $7336 \pm 284$.

\begin{figure}[h]
 \centerline{
\includegraphics[height=10cm]{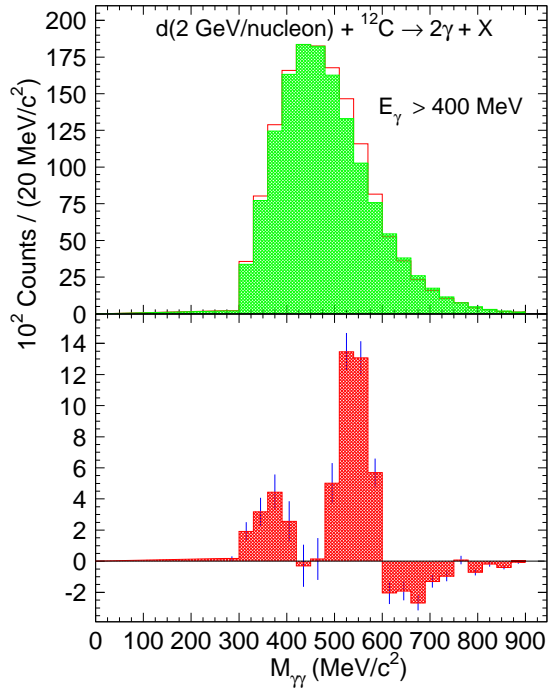}}
  \caption{(color online) The invariant mass spectra of $\gamma\gamma$ pairs
  for the energy selection $E_\gamma>400$ MeV under the selection
criteria $(1)\div (2)$ , $K_{norm} = 0.973$.
 }
 \label{E_sel}
\end{figure}

To elucidate the nature of the detected enhancement, we
investigate the dependence of its position and width on the
opening angle of two photons and on their  energy selection level.
The results demonstrated in Figs. \ref{op_ang} and \ref{E_sel}
show that both maxima survive and are located practically at the
same values of $M_{\gamma\gamma}$.
\begin{figure}[h]
\centerline{
\includegraphics[width=10cm]{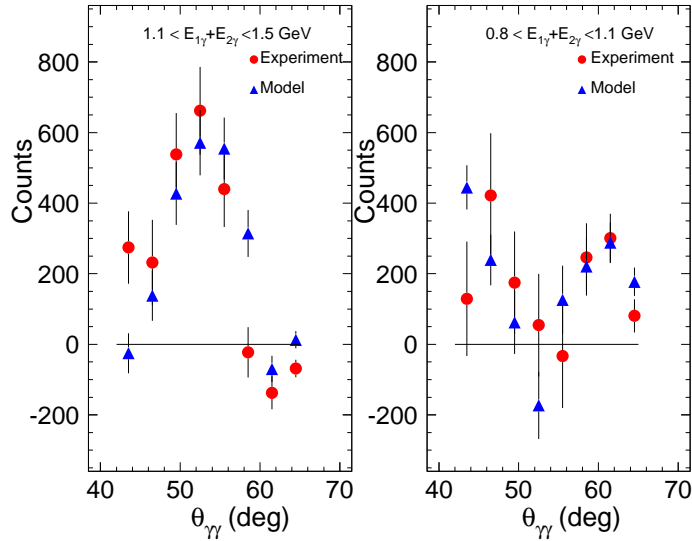}}
  \caption{(color online) Distribution  of the opening angle of
  $\gamma\gamma$ pairs
  in $d$C collisions for the two selections of $(E_{1\gamma}+E_{2\gamma})$.
  Other conditions are the same as in
  Fig.\ref{trig}. Experimental (circles) and simulated (triangles)
  results are normalized to the same number of photon pairs.
 The values of $K_{norm}$ are 0.974 (left) and 0.99 (right).}
\label{angl_dist}
\end{figure}
\begin{figure}[h]
\includegraphics[width=8cm]{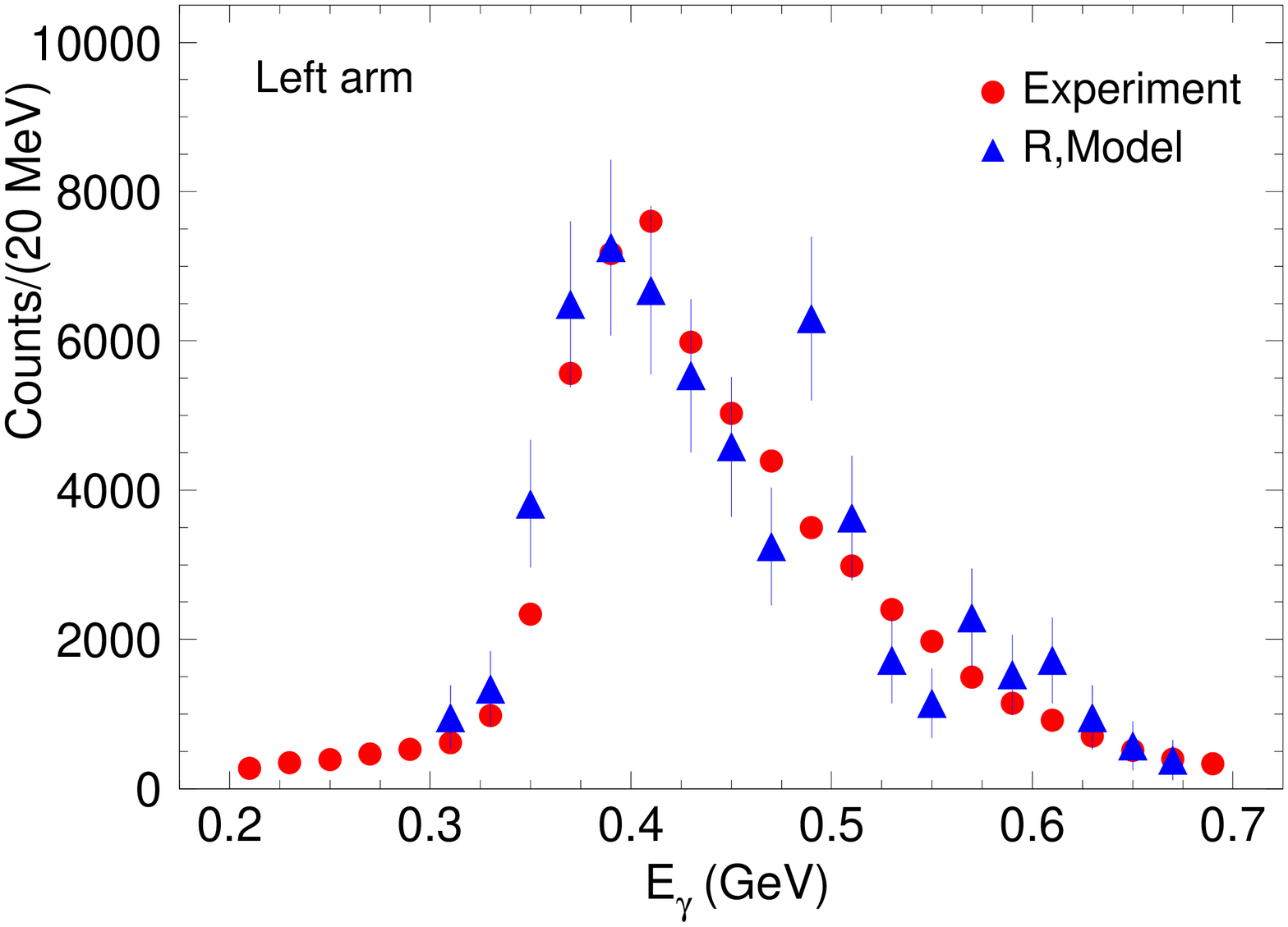}
\includegraphics[width=8cm]{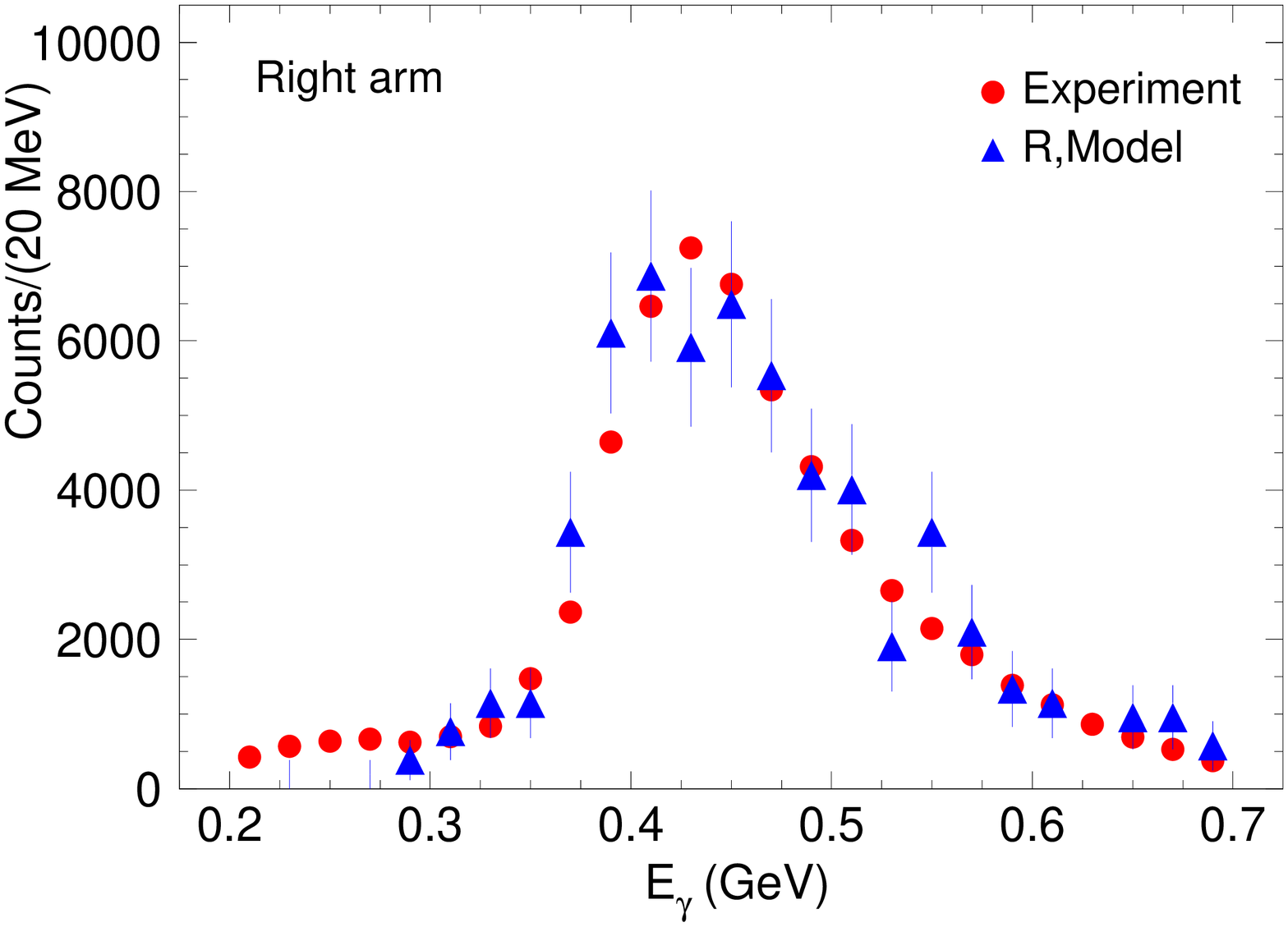}
  \caption{(color online) Photon energy spectra  from $\gamma\gamma$
  pairs in the invariant mass interval $\Delta M_{\gamma\gamma} =0.32\div 0.4$
   GeV. Experimental (circles) and Monte-Carlo simulation (triangles) points calculated with inclusion of the $R$ resonance formation
   are given separately for every spectrometer arm.
   Distributions are normalized to the same total number of events.}
  \label{ph-spectra}
\end{figure}
Pair distributions over the opening angle $\Theta_{\gamma\gamma}$
for two intervals of the sum  of two-photon energy
$0.8<E_{1\gamma}+E_{2\gamma}<1.1$ (right) and
$1.1<E_{1\gamma}+E_{2\gamma}<1.5$ (left) are displayed in Fig.
\ref{angl_dist}. For lower values of the sum energy the two peaks
are seen where the peak at smaller opening angles corresponds to
the $R$ resonance while $\eta$ mesons are emitted at larger angles
around $\Theta_{\gamma\gamma}\sim 60^\circ$. Harder energetic
selection (left panel) leaves only the $\eta$ meson contribution.

As is seen, the result of the changing of observation conditions
is quite robust: the position and width of the observed peak
remain almost unchanged in different intervals of both energies
and opening angles of $\gamma$-quanta, namely, the mean peak
position in the invariant mass distribution  varies under
different conditions in the range $348\div 365$ MeV. The total
number of detected events in the region 270-450 MeV (a summed
number of pairs in the histogram in Fig.\ref{trig}) after the
background subtraction is $2339\pm 340$.

The resonance-like structure  observed in invariant mass di-photon
distributions is not visible in energy photon spectra. As shown in
Fig.\ref{ph-spectra} the $\gamma$ energy spectra near the $R$
resonance are quite smooth, well reproduced by model simulations
and have a very similar shape in both arms. This point testifies
also that there is no instrumental anomaly causing the $R$ peak.

  Because of smallness of the signal-to-background ratio in the
R-resonance mass range, high statistics is needed for observation.
The only most statistically meaningful measurement of the
invariant mass spectra was made by the TAPS-collaboration
\cite{TAPS_CC}. The closest reaction where the $\eta$ production
was studied is CC interactions at the kinetic energy $T_C$ =2 AGeV
\cite{TAPS_CC}. In this experiment $13290\pm 340$ eta mesons were
measured but a resonance structure in the range of
$M_{\gamma\gamma}=300\div 400$ MeV has not been recorded. If the
$R/\eta$ ratio is assumed to be the same as in the case of $d$C
collisions, one may expect about $1800\cdot (\epsilon_R
/\epsilon_\eta)$ of the detected $R\rightarrow \gamma\gamma$
events, where $\epsilon_R$ and $\epsilon_\eta$ are the detection
and selection efficiency for $R$ and $\eta$, respectively. Taking
into account the systematical uncertainties, this estimated value
is in the limits of $(500 \div 3000)\cdot (\epsilon_R
/\epsilon_\eta)$ of the detected
 $R\rightarrow \gamma\gamma $ events (see below, formula
(\ref{eq36})). The total number of detected events in this range
of two-photon invariant masses is about $6.\cdot 10^5$
\cite{TAPS_CC}, so roughly the resonance structure should be
observed at the level of $\sim 1800\cdot (\epsilon_R
/\epsilon_\eta) \pm 800$. To resolve this structure general
statistics should be increased by an order of magnitude, at least.
Note that these two experimental setups cover different rapidity
regions, which was not taken into account in our estimate. As
compared to the TAPS, the PHOTON-2 setup has a smaller angular
acceptance but a better signal-to-background ratio. Thus, there is
no discrepancy between our result and observation of no resonance
structure by the TAPS \cite{TAPS_CC} in the reaction close in the
energy and mass numbers.

 An indication to a possible resonance structure in pp-collisions
at the energies 1.36 and 1.2 GeV was obtained by the CELSIUS-WASA
collaboration \cite{Bea05}. However statistics in the invariant
mass range 250-350 MeV is low:  only about 200 $\gamma\gamma$
pairs without background subtraction was found.

\section{Wavelet analysis}

Here we shall try to identify essential structures in the measured
$M_{\gamma\gamma}$ spectra without the background subtraction. A
conventional method for such analysis is a wavelet transformation
which is known as an efficient multiscale technique  to reduce the
presence of statistical noise and then extract physical parameters
from the obtained smoothed form \cite{H95,T95}. The
one-dimensional wavelet transform (WT) of a signal $f(x)$ has a
biparametric form. This allows WT to overcome the main
shortcomings of the Fourier transform such as nonlocality,
infinite support  and necessity of a broad band of frequencies to
decompose even a short signal. The wavelet transformation changes
the decomposition basis into functions which are compact into a
time/space and frequency domain. The WT with the wavelet function
$\psi$ of the function $f(x)$ is defined by convolution as
\begin{eqnarray}
W_\psi (a, b)f=\frac{1}{\sqrt{C_\psi}}\int\limits_{-\infty}^\infty
\frac{1}{\sqrt{|a|}}\psi\left(\frac{b-x}{a}\right)f(x)\;dx
\label{cwt}
\end{eqnarray}
with the normalization constant
$$C_\psi=2\pi\int\limits_{-\infty}^\infty\frac{|\tilde{\psi}
(\omega)|^2}{|\omega|}\;d\omega\,\,<\,\,\infty,$$ where
$\tilde{\psi}(\omega)$ is the Fourier transform of the wavelet
$\psi(x)$. A scale parameter $a$ characterizes the dilatation and
$b$ is the translation in time or space. In this respect the
wavelet function $\psi(t)$ is a sort of a "window function" with a
non-constant window width: high-frequency events are narrow (due
to the factor $1/a$),  while low-frequency wavelets are broader.
The inverse transform is given by the formula
\begin{equation}
f(t) = C_\psi^{-1}\int\int \psi\left( \frac{t-b}{a} \right)
W_\psi(a,b) \ \frac{da \ db}{a^2}~. \label{icwt}
\end{equation}

The wavelet $\psi$ exists if $C_\psi < \infty$. It holds, in
particular, when the first $(n-1)$ moments are equal to zero
\begin{equation}
 \int\limits_{-\infty}^\infty |x|^m \psi(x)dx=0,
\quad 0\le m< n .
 \label{moment}
\end{equation}

Due to freedom in the choice of the wavelet function $\psi$, many
different wavelets were invented \cite{{Daubec},{gwave}}. We
consider here only continuous Wavelets with Vanishing Moments
(WVM) (see {it Appendix A}). The WVM family is called so because
condition (\ref{moment}) always holds for it. One of WVM families
is a set of {\it Gaussian wavelets} (GW) which are normalized
derivatives of the Gauss function
\begin{equation}
 g(x;A,x_0)=A\exp\left(-\frac{(x-x_0)^2}{2\sigma^2}\right).
\label{oneD}
\end{equation}

 In some cases continuous wavelets are more suitable to evaluate
peak parameters. One of these cases arises when a peak in question
has a Gaussian shape (\ref{oneD}). This makes it possible to use
very simple analytical expressions, eq.(\ref{gausswavecoef}), in
the continuous Gaussian wavelet transform for Gaussian peaks. It
gives us a remarkable advantage to calculate the peak parameters
directly in the wavelet domain instead of the time/space domain
{\it without using the inversion}. Moreover, in real cases, when
our signal shape is close to a Gaussian one and is considerably
contaminated by an additive noise and, in addition, is distorted
by binning to be input to computer, one can also use the
remarkable robustness of Gaussian wavelet filtering, as proved in
\cite{gwave}.

Wavelet ability to separate signal components with different
frequencies and positions has attracted many physicists to use
both discrete and continuous wavelets \cite{dremin, astafeva}.
Usually the conventional filtering approach is applied: a signal
transformed by a wavelet undergoes an appropriate thresholding and
then is restored by the inverse transform. The image of the
wavelet spectrum is used to obtain rough parameter estimations of
wanted peaks of invariant mass spectra, as in \cite{perm} where
the Mexican hat wavelet was used. In our paper, we apply the
family of GW to look for peaks in question having a Gaussian shape
(\ref{oneD}).

\begin{figure}[h]
 \includegraphics[width=8cm]{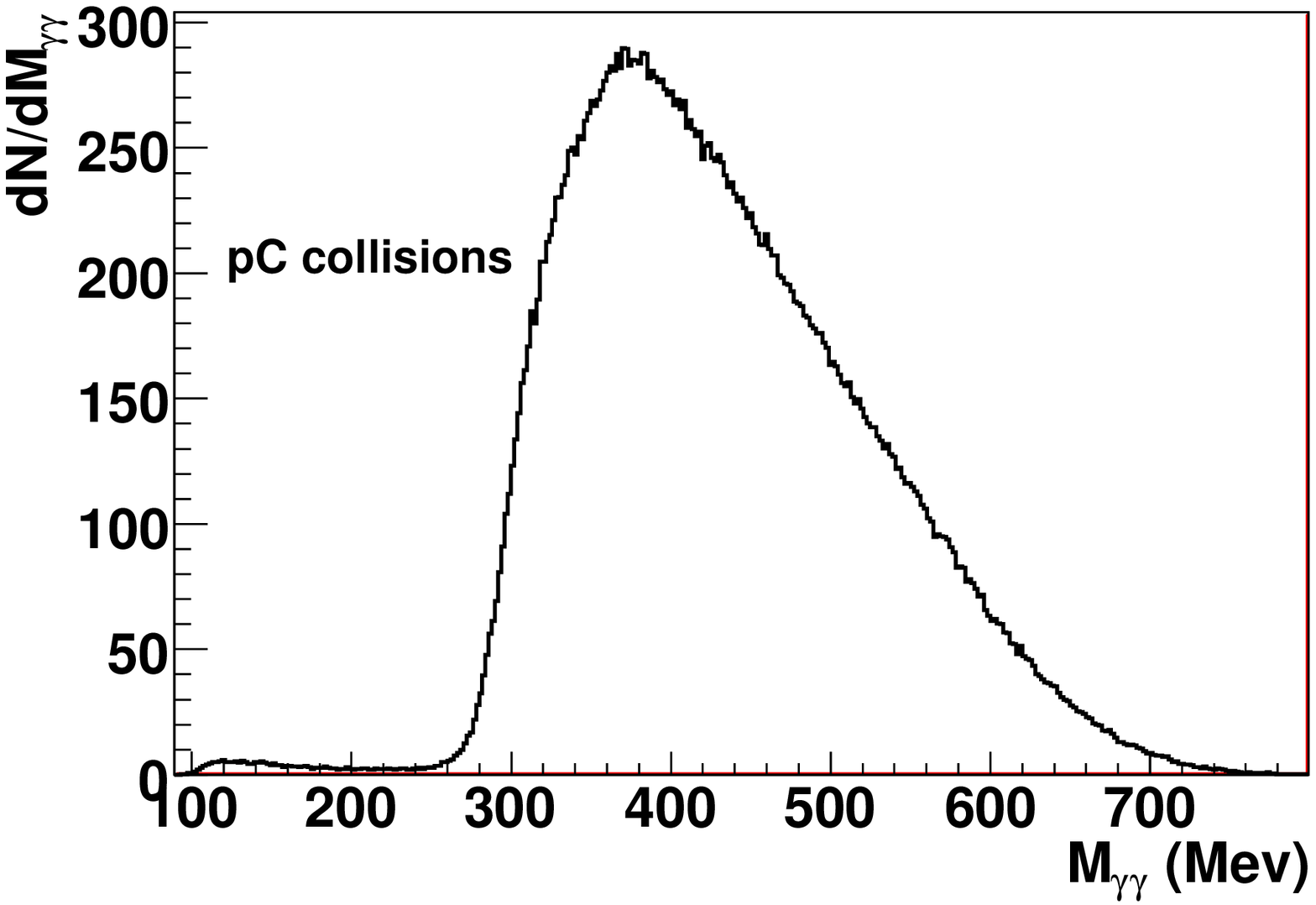}
 \includegraphics[width=8cm]{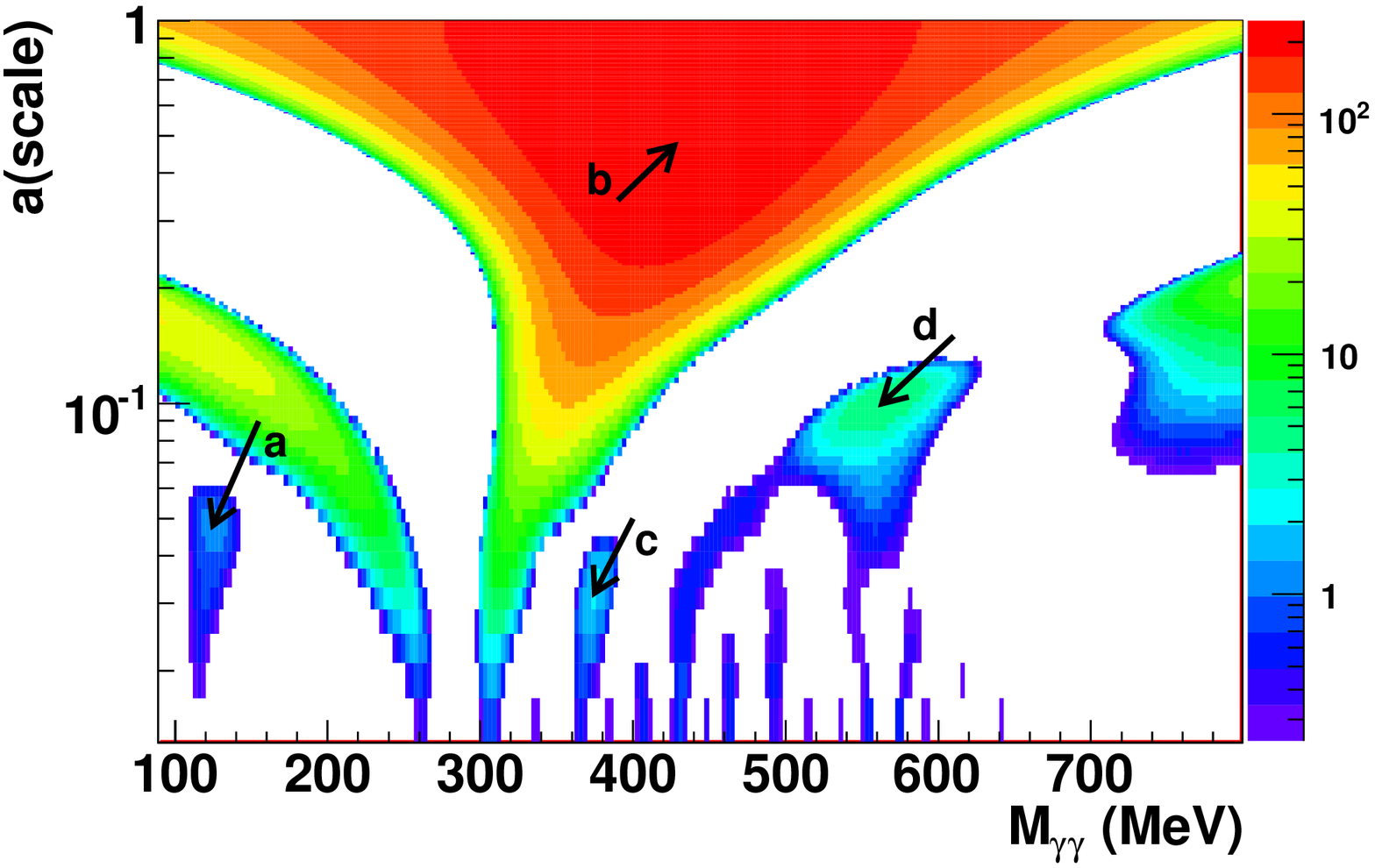}
 \caption{(color online) The invariant mass distribution of $\gamma\gamma$
  pairs
(top) and the biparametric distribution of the GW of the 8-th
order (bottom) for $p$C interactions. These events are selected
under the same conditions as in Fig.\ref{trig} but the
distribution is obtained with an additional condition for photon
energies $E_{\gamma 1}/E_{\gamma 2}>0.8$ and binning in 2 MeV.
 }
  \label{wlpC}
\end{figure}
\begin{figure}[h]
  \includegraphics[width=8cm]{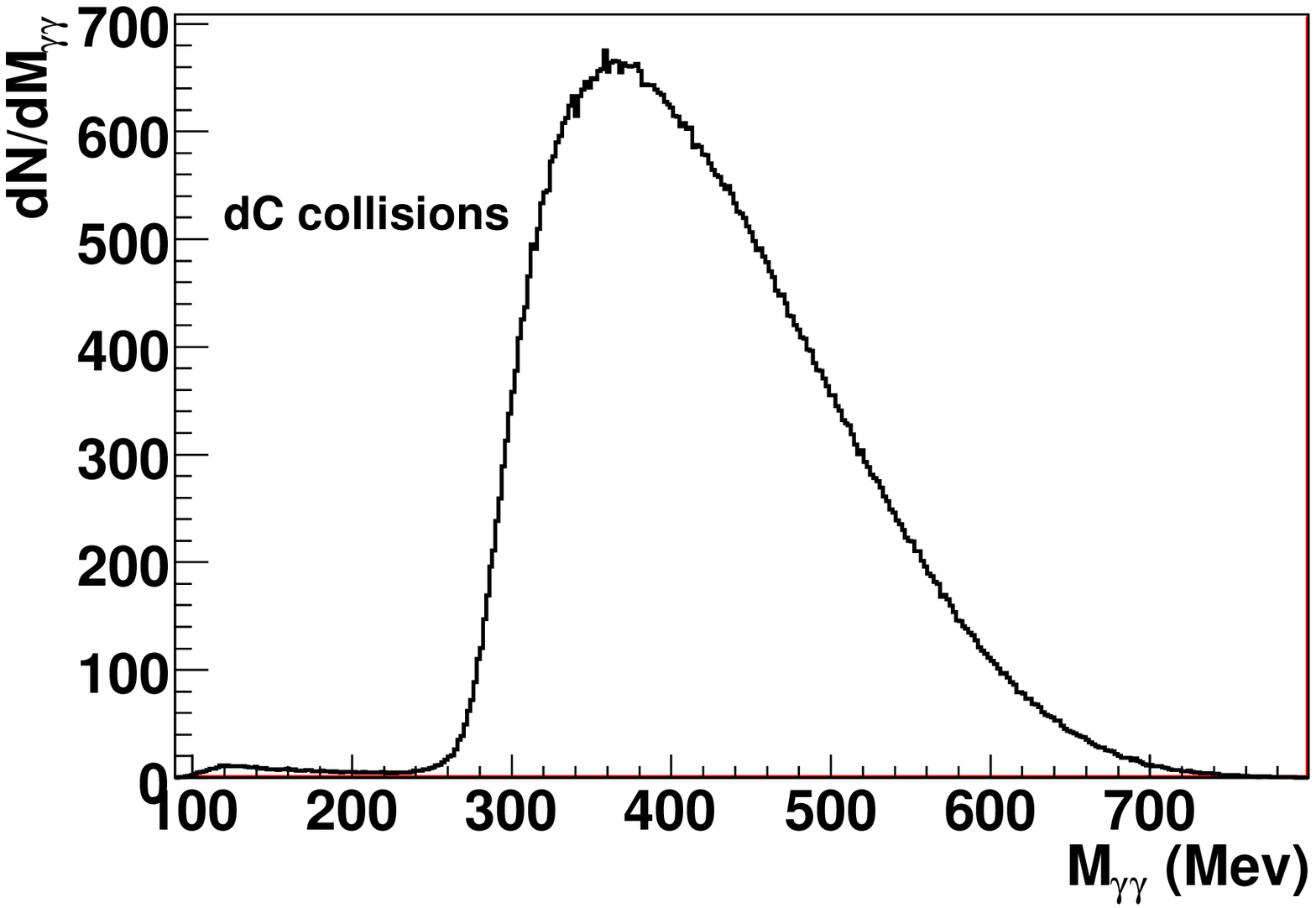}
\includegraphics[width=8cm]{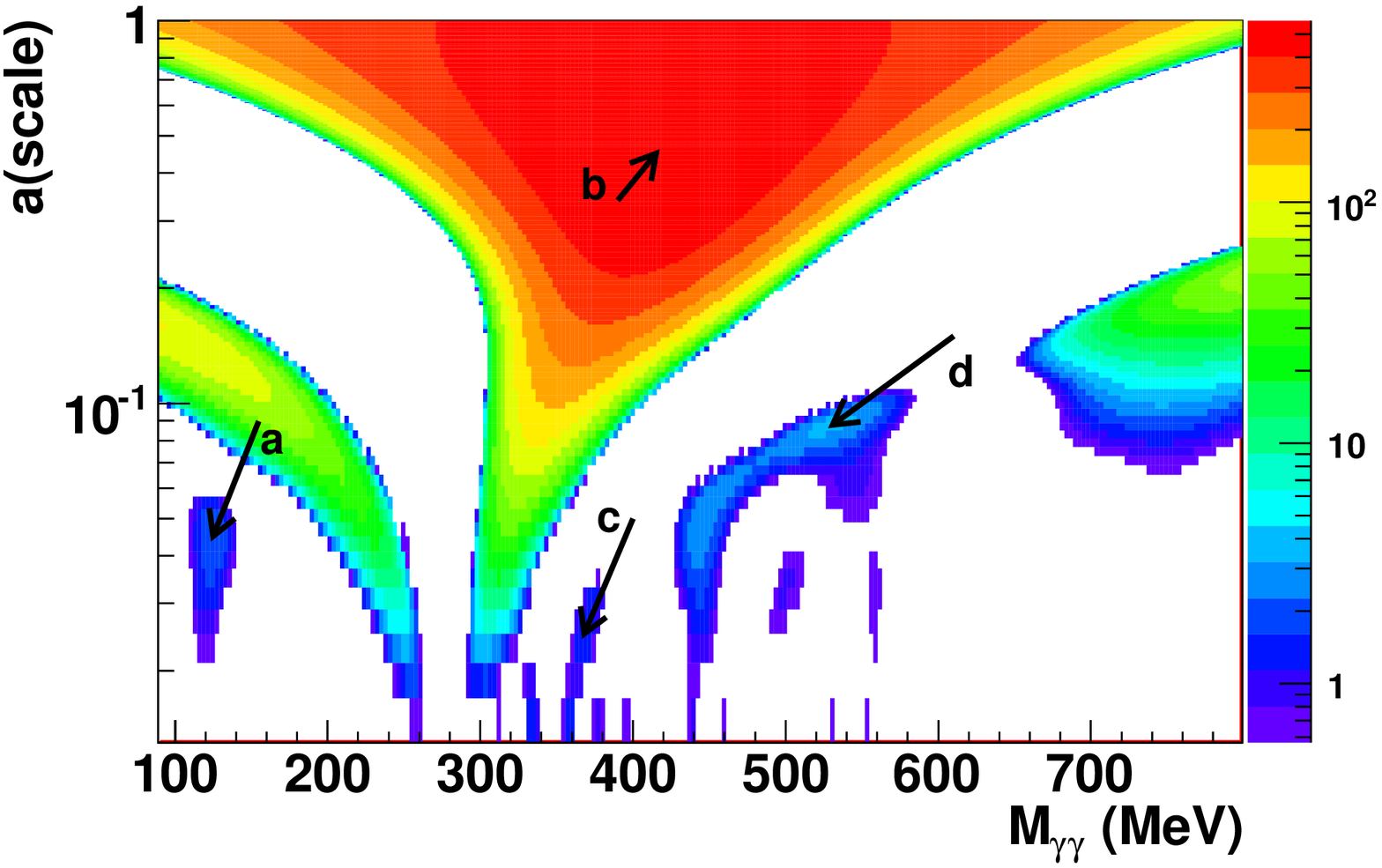}
  \caption{(color online)
The same as in Fig.\ref{wlpC} but for $d$C collisions. }
  \label{wldC}
\end{figure}

The main idea of our approach is to transform the signal $f(x)$ to
the space of the corresponding wavelet and look there for a local
biparametric area surrounding the wavelet image of our peak in
question, drawing down all other details of the signal image,
concerning noise, binning effects, and background pedestal. A
particular example of the $G_2(a,b)$ transform and other details
of this analysis are given in {\it Appendix A}.

The initial signal $f(x)$, {\it i.e.} the $M_{\gamma\gamma}$
distribution including the background is presented in the upper
panels of Figs.\ref{wlpC} and \ref{wldC}. Here an additional
condition on photon energies $E_{\gamma 1}/E_{\gamma 2}>0.8$
(where $E_{\gamma 1}$ and $E_{\gamma 2}$ are the smallest and the
largest energy in the given photon pair, respectively) was applied
to data to improve the signal-to-background ratio and avoid
double-humped structure which will require higher orders of the
expansion. The distributions look quite smooth due to the choice
of a smaller step in $M_{\gamma\gamma}$ which is needed to provide
more points for the wavelet analysis.

The wavelet transformation results are presented in the lower
panels of Figs. \ref{wlpC} and \ref{wldC} for $p$C and $d$C,
respectively. Assuming symmetric signal we use the WVM of the 8-th
order  to separate noise and see a peak structure. Attempts to
apply wavelets of lower than 8-th order give worse results,
perhaps, because of rather ragged signals. The arrows show an
approximate location of the identified peaks. Due to  the trigger
condition, the distribution maximum $(a)$ of photons from the
$\pi^0$ decay is shifted to  $M_{\gamma\gamma} \sim 125$ MeV from
the expected 135 MeV. It is in agreement with the initial
distribution shown in the same figures. A huge peak from the
background $(b)$ dominates and its shape is quite close to the
Gaussian. A photon peak from the $\eta$ decay, $(d)$, is seen
quite distinctly at the proper place but its intensity is
suppressed due to an additional condition $E_{\gamma 1}/E_{\gamma
2}>0.8$. In the domain of interest there is some enhancement near
$M_{\gamma\gamma}\sim 370$ MeV marked in the figures by $(c)$. In
the biparametric representation this spot has not a circle shape
because even in the case of a coarse binning it is not well
approximated by the Gaussian (see Tabl.\ref{tab2}). In this
respect, the  $(c)$-peak in $p$C seems to be more pronounced than
the appropriate one in $d$C but it follows from lower statistics
in the $p$C case where a separate point may be better approximated
by the Gaussian. Note that statistics in these cases differs by
the factor of more than 3. The WVM analysis still reveals one more
weak $(c)$-peak at higher $M_{\gamma\gamma}$. This is not very
surprising since in coarse binning (see Fig.\ref{f3}) they were
blurred but they are seen at a more strict selection (cf.
Figs.\ref{trig} $\div$ \ref{E_sel}).

Therefore, the presented results of the continuous wavelet
analysis with vanishing moments confirm the finding of a peak at
$M_{\gamma\gamma}\sim (2-3)m_\pi$  in the $\gamma\gamma$ invariant
mass distribution obtained within the standard method with the
subtraction of the background from mixing events.

\section{Data simulation }
\subsection{About the model}

To simulate $p$C and $d$C reactions we use a transport code. At
high energies it is the Quark-Gluon String Model (QGSM)
\cite{QGSM} and at the energy of a few GeV the string dynamics is
reduced to the earlier developed Dubna Cascade Model (DCM)
\cite{Dub_casc} with upgrade of elementary cross sections involved
\cite{LAQGSM}.

The DCM divides the collision into three stages, well separated in
time. During the first initial stage an intranuclear cascade
develops, primary particles can scatter and secondary particles
can re-scatter several times prior to their absorption or escape
from the nucleus. At the end of this step the coalescence model is
used to localize $d$, $t$,$^{3}$He, and $^{4}$He particles from
nucleons found inside spheres with well-defined radii in
configuration space and momentum space. The emission of cascade
particles determines a particle-hole configuration, i.e., Z,A, and
excitation energy that is taken as the starting point for the
second, pre-equilibrium stage of the reaction, described according
to the cascade exciton model \cite{exciton}. Some pre-equilibrium
particles may be emitted and this leads to a lower excitation of
the thermalized residual nuclei. In the third, final
evaporation/fission stage of the reaction, the de-excitation of
the residue is described with the evaporation model. The last two
stages are important for triggering the events. All components
contribute normally to the final spectra of particles and light
fragments; low-energy evaporated photons are not included into
subsequent analysis. For relativistic energies the cascade part of
the DCM is replaced by the refined cascade model, which is a
version of the quark-gluon string model (QGSM) developed in
\cite{Toneev90} and extended to intermediate energies in
\cite{Amelin90}. The description of the mean-field evolution is
simplified in the DCM in the sense that the shape of the scalar
nuclear potential, defined by the local Thomas-Fermi
approximation, remains the same throughout the collision. Only the
potential depth changes in time, according to the number of
knocked-out nucleons. This frozen mean-field approximation allows
us to take into account the nuclear binding energies and the Pauli
exclusion principle, as well as to estimate the excitation energy
of the residual nucleus by counting the excited particle-hole
states. This approximation is usually considered to work
particularly well for hadron-nucleus collisions.

The following $\gamma$-decay channels are taken into account: the
direct decays of $\pi^0,  \eta, \eta'$ hadrons into two
$\gamma$'s,
$\omega\to\pi^0\gamma$, $\Delta \to N\gamma $ and the Dalitz decay
of $\eta\to \pi^+\pi^-\gamma$, $\eta\to \gamma e^+ +e^-$ and
$\pi^0\to \gamma e^+ +e^-$, the $\eta'\to \rho^0+\gamma$, the
$\Sigma \to \Lambda+\gamma$, the $\pi N$ and $NN$-bremsstrahlung.
One should note that in accordance with the recent HADES
data~\cite{HADES}, the $pn$-bremsstrahlung turned out to be higher
by a factor of about 5 than a standard estimate and weakly depends
on the  energy. This finding, being in agreement with the recent
result of Ref.\cite{Copt}, allowed one to resolve the old DLS
puzzle \cite{Brat07}. This enhancement factor is included in our
calculations.
\begin{figure}[h]
\centerline{\includegraphics[width=8cm]{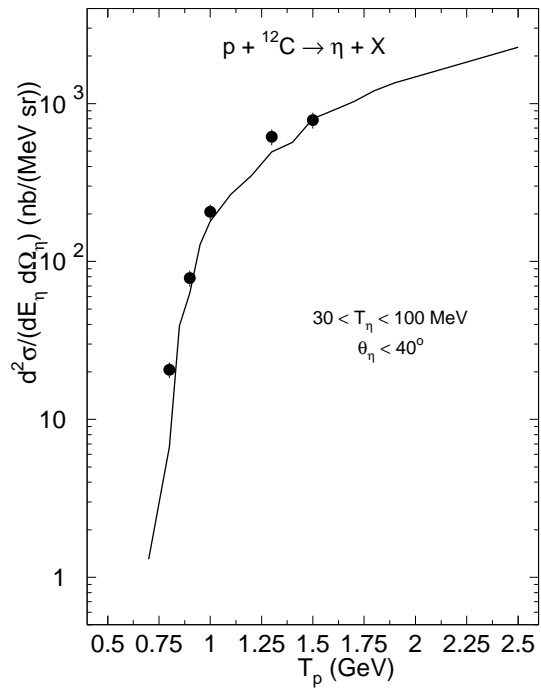}
 }
  \caption{
The proton energy dependence of the double differential cross
section for the $\eta$ production in $p$C collisions. Experimental
points are from \cite{Ch92}.
  }
\label{pceta1}
\end{figure}
\begin{figure}[h]
\centerline{\includegraphics[width=14cm]{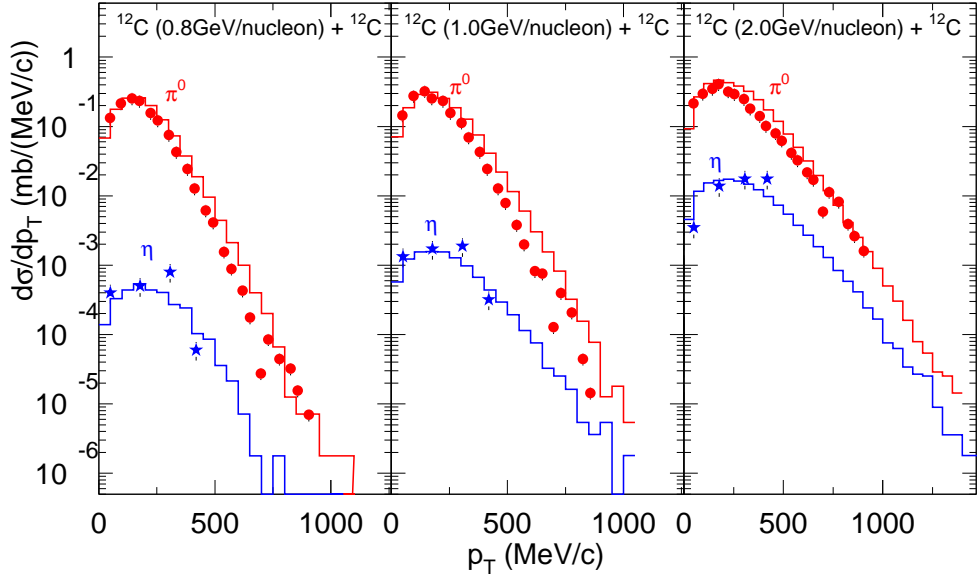}}
 \caption{(color online)
Transverse momentum distributions of $\pi^0$ and $\eta$ in the
middle rapidity range from $CC$ collisions at different energies.
Experimental points are from the TAPS Collaboration
\cite{TAPS_CC}.
  }
  \label{pt_pi0eta}
\end{figure}

As a model test, in Fig.\ref{pceta1} the excitation function for
the $\eta$ production is shown for the $p$C collisions. The model
describes correctly a fast increase of the $\eta$ yield near the
threshold where the cross section is changed by two orders of
magnitude.

The transverse momentum distributions at the mid-rapidity are
presented in Fig.\ref{pt_pi0eta} for $\pi^0$ and $\eta$ produced
in $CC$ collisions at three bombarding energies. The model results
are in good agreement with the TAPS experiment \cite{TAPS_CC} for
both neutral pions and eta mesons. So this gives us some
justification for application of our model to analyze neutral
particle production in the reactions considered.

\subsection{Analysis of the obtained data}
\begin{figure}
 \centerline{\includegraphics[width=8cm]{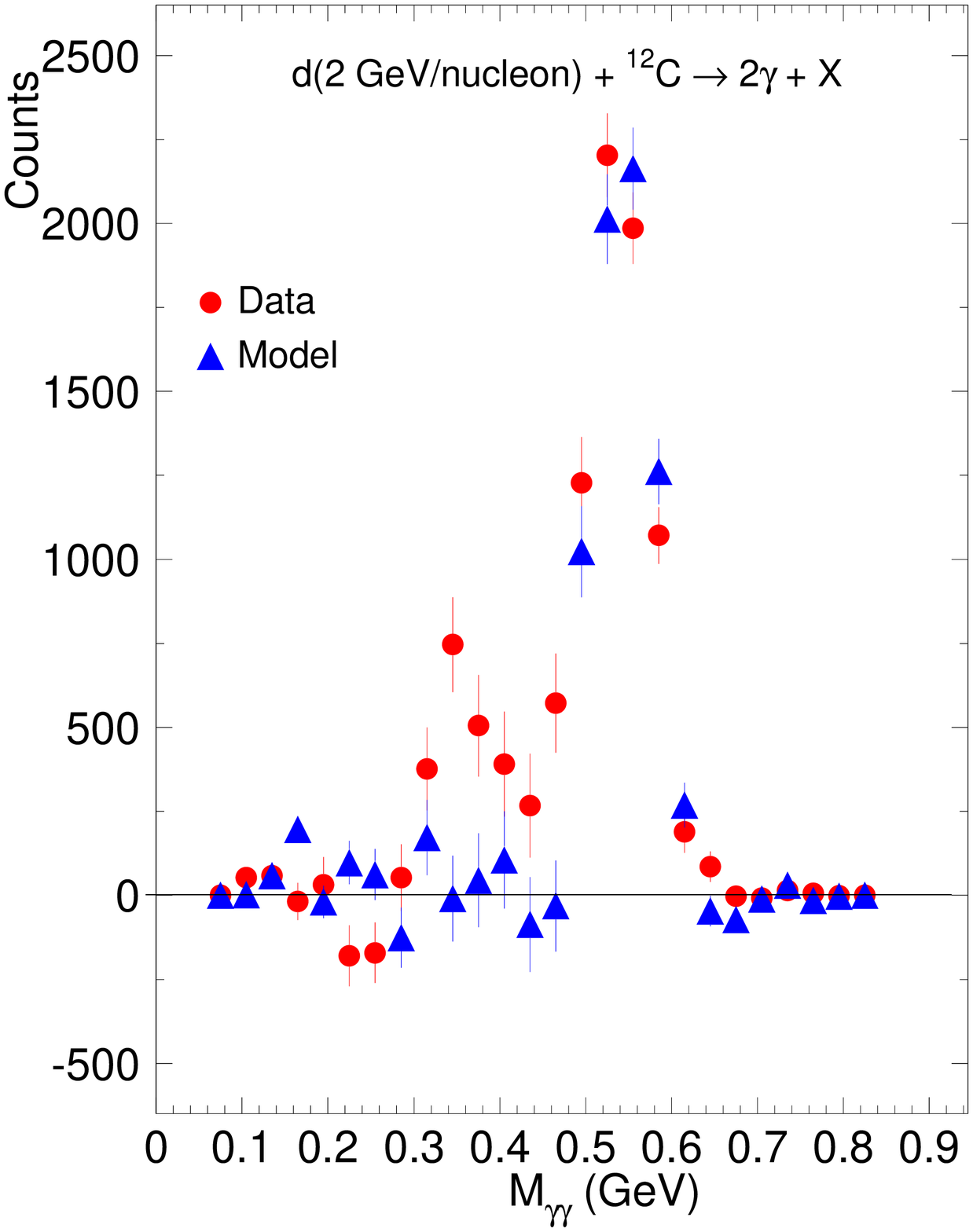}
 \includegraphics[width=8cm]{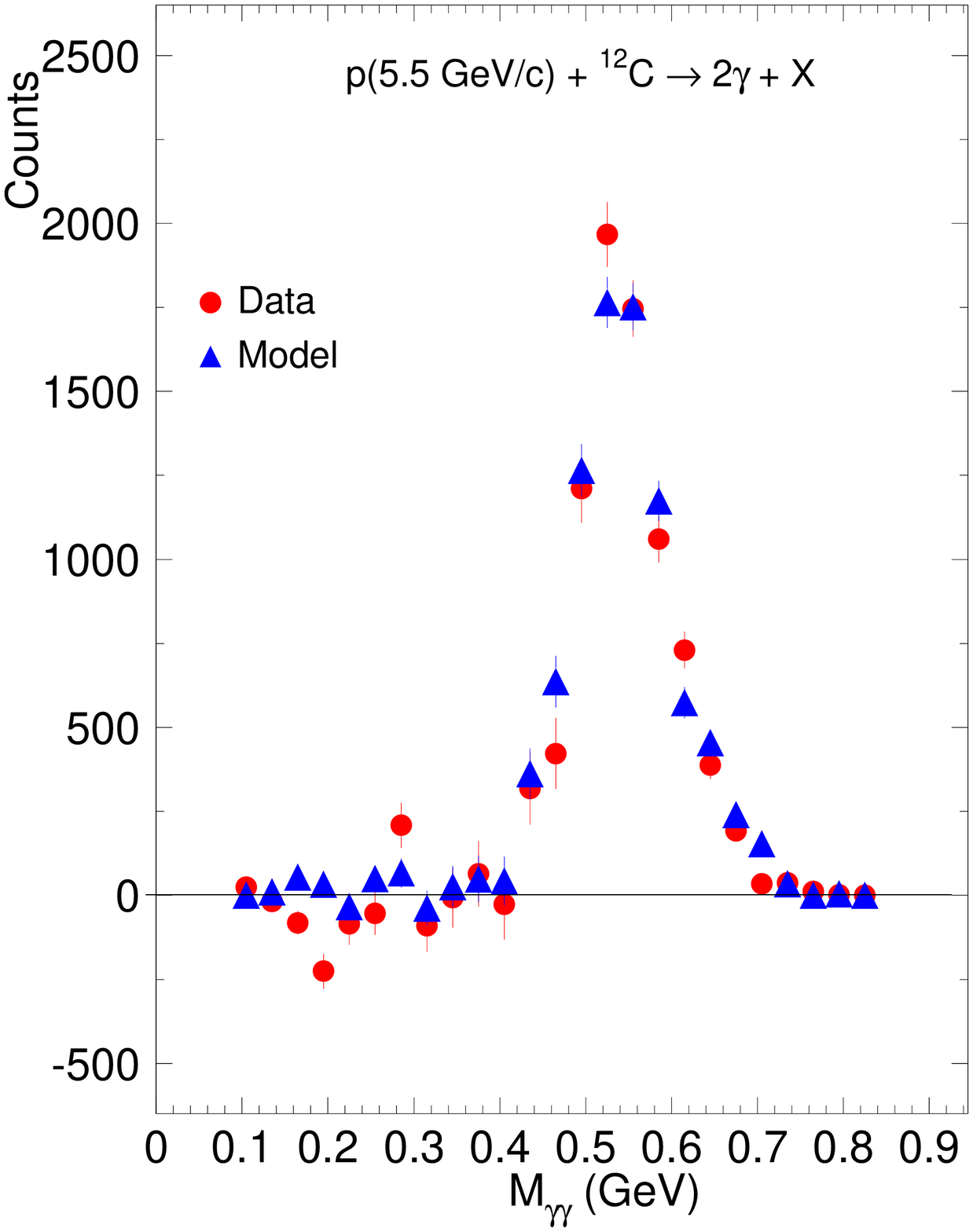}}
 \caption{(color online)The invariant mass distributions of $\gamma\gamma$ pairs
  from the $d$C (left) and $p$C (right) reactions after background
  subtraction. Both experimental (circles) and simulated (triangles)
  points are obtained under the same PHOTON-2 conditions.
  }
  \label{dC-backgr}
\end{figure}

The model described above is implemented for describing the
measured distributions with careful simulations of experimental
acceptance. The total statistics of simulated events amounts to
about $10^9$ here and in every case below. As is seen from
Fig.\ref{dC-backgr}, the model reproduces quite accurately the
observed $\eta$ peak in the invariant mass distribution of
$\gamma$ pairs but there is no enhancement in the region of the
$R$-resonance.

\begin{figure}
\centerline{\includegraphics[width=8cm]{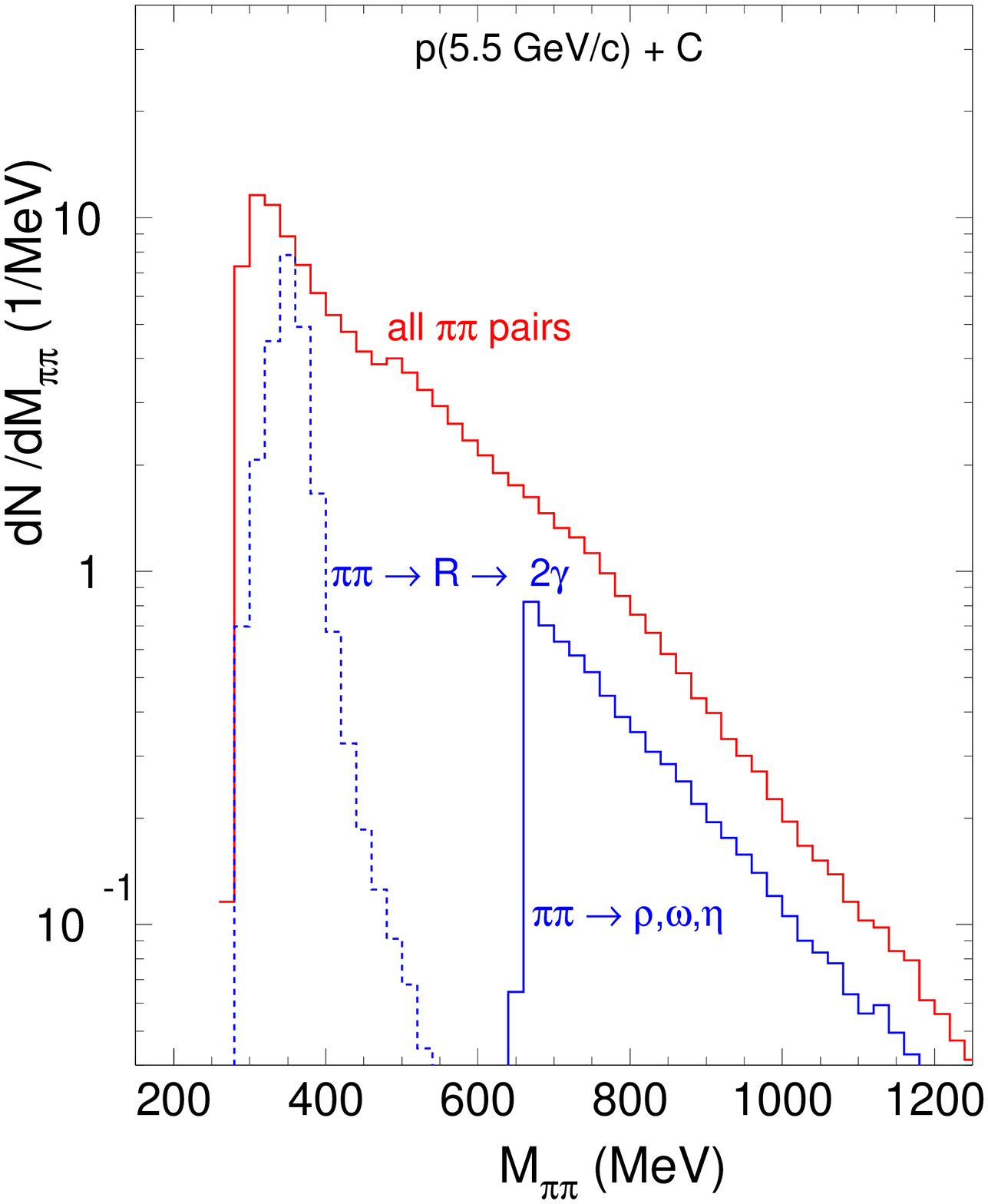}
\includegraphics[width=8cm]{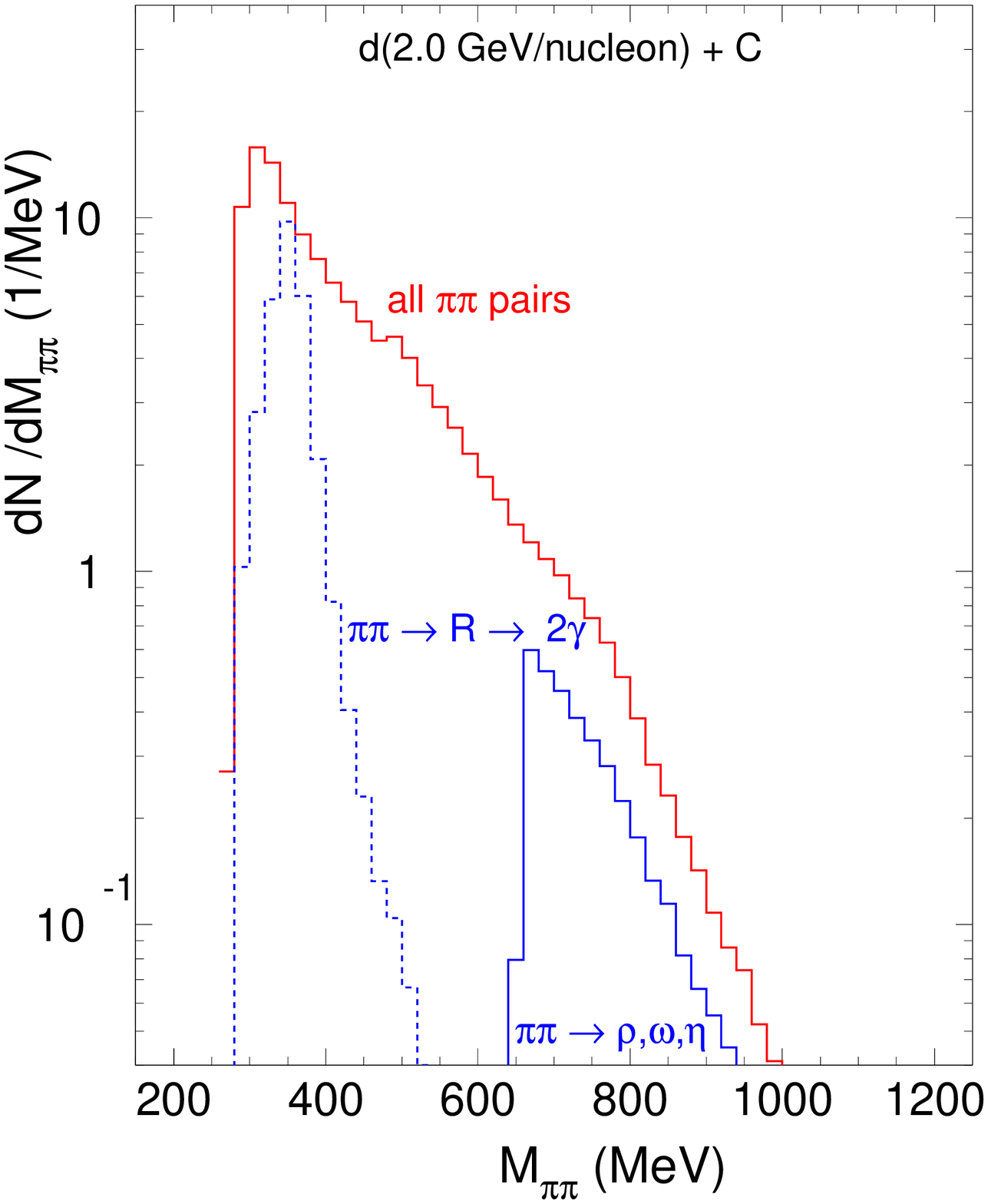} }
 \caption{
 Invariant mass distribution of pion pairs from $\pi\pi$
interactions in $p$C (left) and $d$C (right) collisions. The
contributions resulting in the $R$-resonance and formation of
$\rho, \omega, \eta'$ mesons are shown separately.
  }
  \label{pCdCM}
\end{figure}
The mass of the $R$ resonance is slightly above $2m_\pi$. As was
noted, there is no particularities in the phase shifts  for the
$\pi\pi$ scattering. We remind that according to the PDG table
\cite{PDT06},  the closest-in-mass hadron in this energy range is
the $f_0(600)$-meson (or the $\sigma$-meson) with the extended
mass and very large width (see eq.(\ref{PDT})); however, the
recent analysis gives a more defined $\sigma$ mass and a more
narrow width (see eq.(\ref{CCL})).

To see whether such a resonance structure would be created due to
a nuclear interaction and can survive in the strict experimental
conditions, we artificially simulate production of the
$R$-resonance and follow its fate in the course of a nuclear
collision. It is assumed that the putative $R$ resonance can be
created in every $\pi^+\pi^-$ or $\pi^0\pi^0$ interaction if its
invariant mass $M_{\pi\pi}$ satisfies the Gaussian distribution
with the observed parameters (see Tabl.\ref{tab2}). The
$M_{\pi\pi}$ distributions for proton- and deuteron-induced
reactions is presented in Fig.\ref{pCdCM}. These distributions are
rather wide with a pronounced peak in the region of (2-3)$m_\pi$.
The fraction of interactions identified with the formation of the
$R$-resonance is slightly above  20 $\%$ of all $\pi\pi$
collisions. Interactions with $M_{\pi\pi}\sim 650$ MeV resulting
in  heavy mesons $\rho, \omega, \eta$ come to about  5 $\%$.
\begin{figure}
\centerline{\includegraphics[width=8cm]{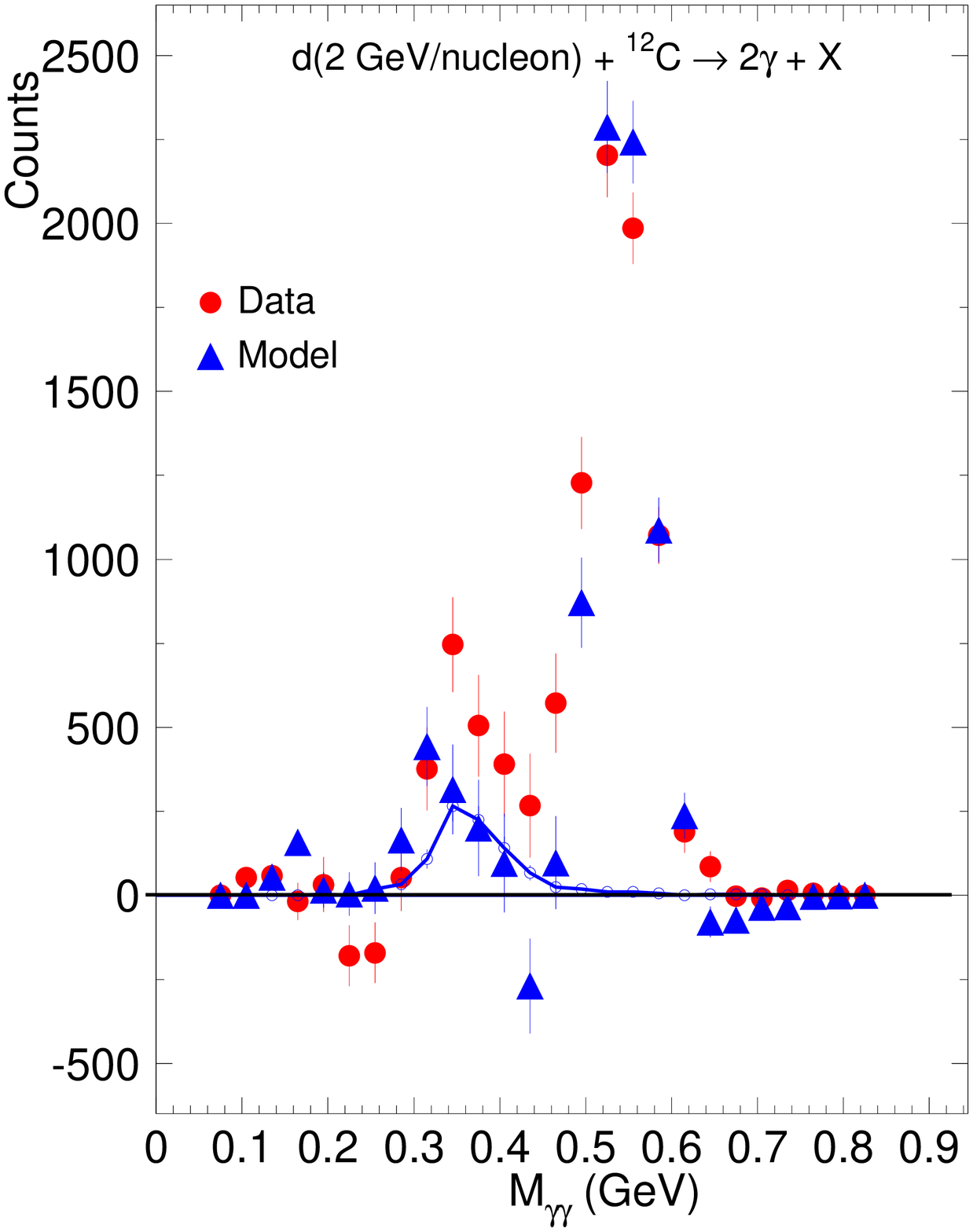}
\includegraphics[width=8cm]{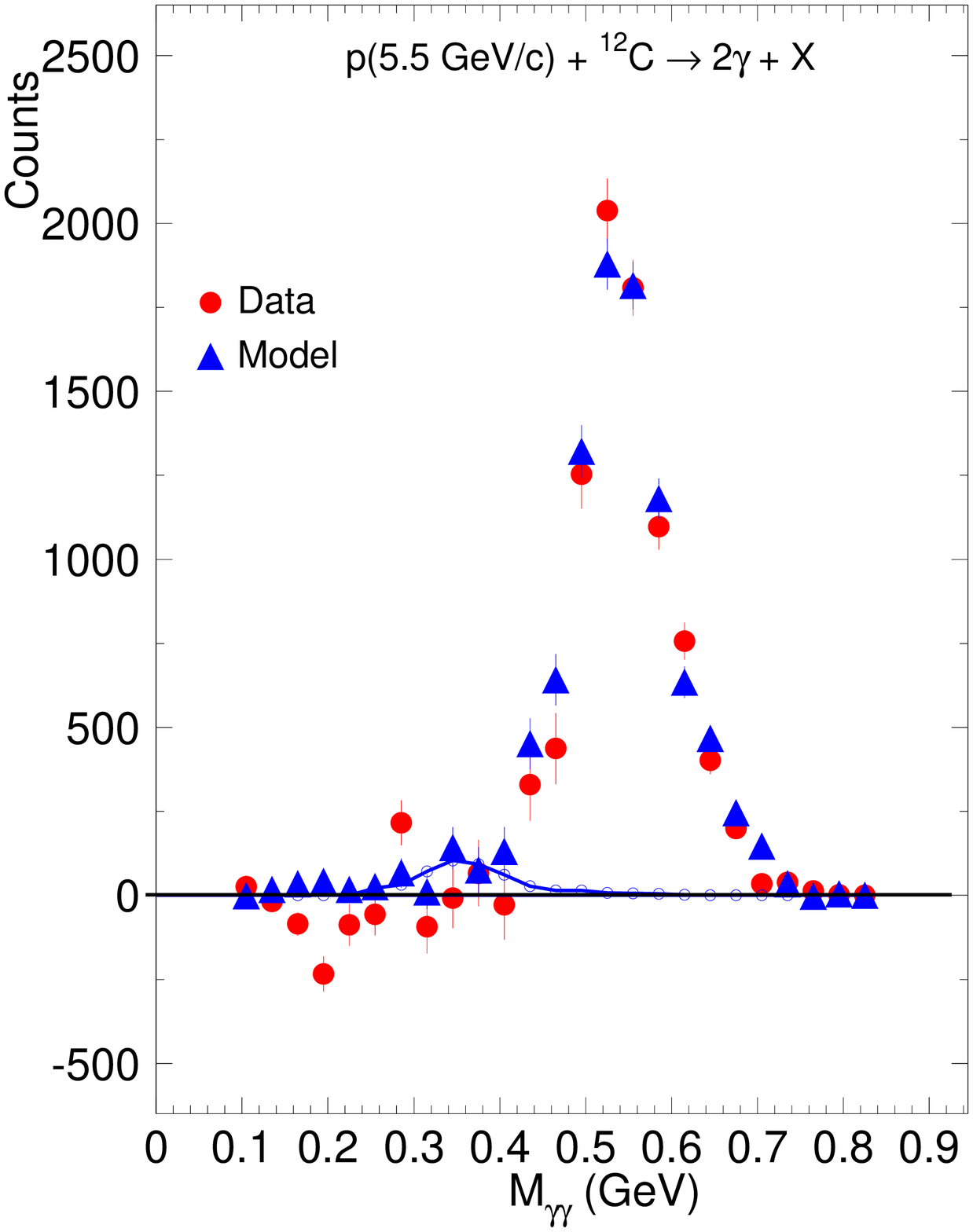}}
  \caption{(color online) Invariant mass distributions of $\gamma\gamma$ pairs
from the $d$C and $p$C reactions after background subtraction.
Both experimental (circles) and simulated (triangles) points are
obtained under the same conditions. The contribution of photons
from the $R$ decay is shown by the solid line.
 }
 \label{pCdCR}
\end{figure}

If the $R$ resonance has been formed, it is assumed to decay only
in two photons. This scheme can be easily realized by the Monte
Carlo method within our transport model. The two-photon invariant
mass spectra calculated with the inclusion of the possible $R$
production are compared with experiment in Fig.\ref{pCdCR}.
Indeed, the essential part of $\gamma\gamma$ pairs survives
through the strict selected rules and can explain $(20-30)\%$ of
the enhancement in the case of $d$C collisions. For $p$C
collisions the $R$ contribution is smaller but in agreement with
experimental points. Nevertheless, one should be careful in taking
too seriously the  absolute values of the $\gamma\gamma$ pair
yields estimated from the $R$ decay. They are obtained under
extreme assumption that all $R$ resonances do decay via the
two-photon channel, {\it i.e.} $\Gamma_R= \Gamma_{\gamma\gamma}$.
However, the scalar resonance (like the $\sigma$ meson) decays
mainly into two pions, $\Gamma_R=\Gamma_{\pi\pi}$, and the
electromagnetic decay is strongly suppressed. The two-photon decay
is dominating ($\Gamma_R=\Gamma_{\gamma\gamma}$) only if the $R$
mass is below the two-pion threshold. So the proposed mechanism
allows one to consider properly the kinematics of the
$\gamma\gamma$ pairs and the role of acceptance, but it is not
able  to describe the absolute $R$ yield which should be by a few
orders of magnitude lower than that presented in Fig.\ref{pCdCR}.
\begin{figure}[h]
\centerline{\includegraphics[width=8cm]{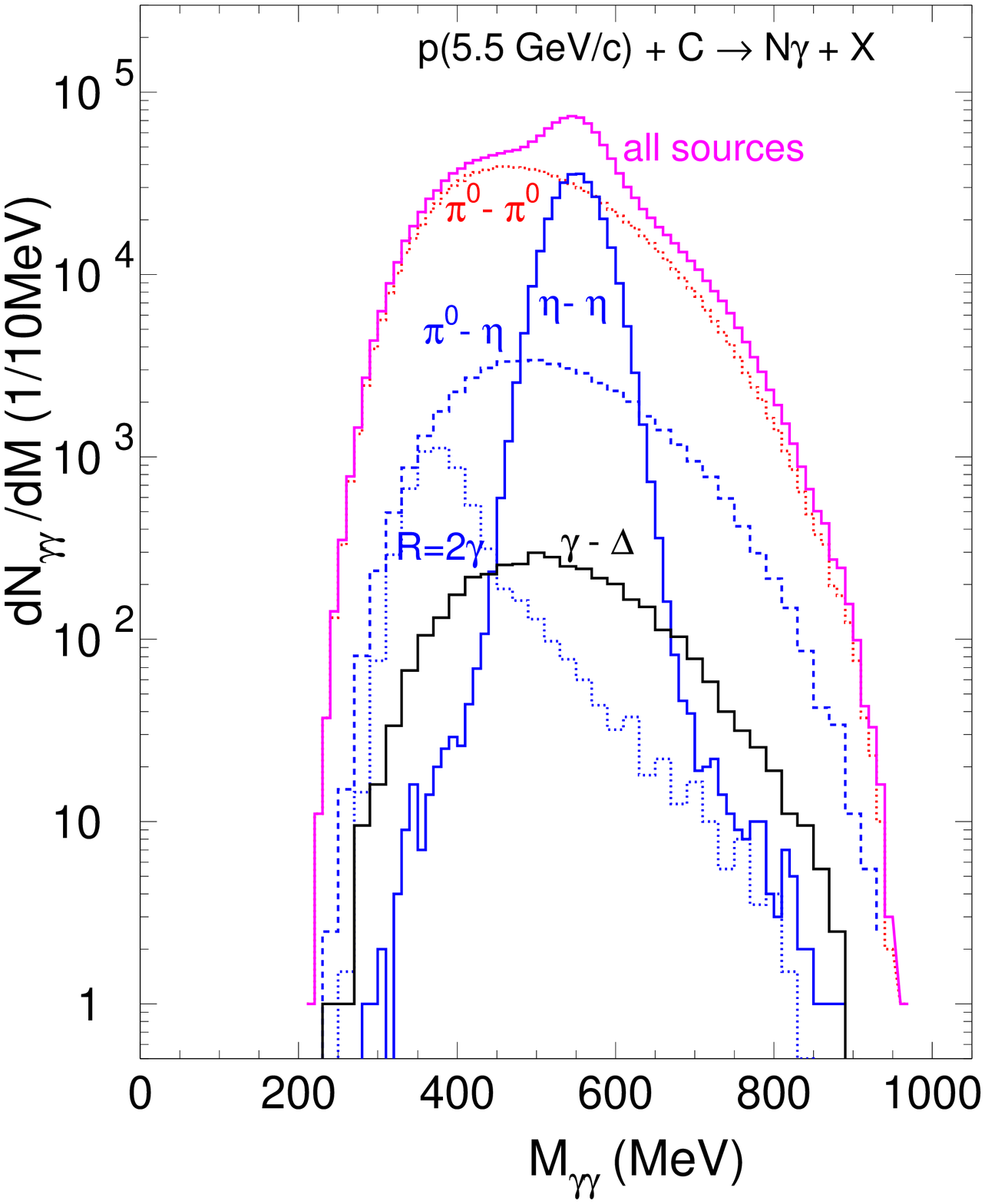}
  \includegraphics[width=8cm]{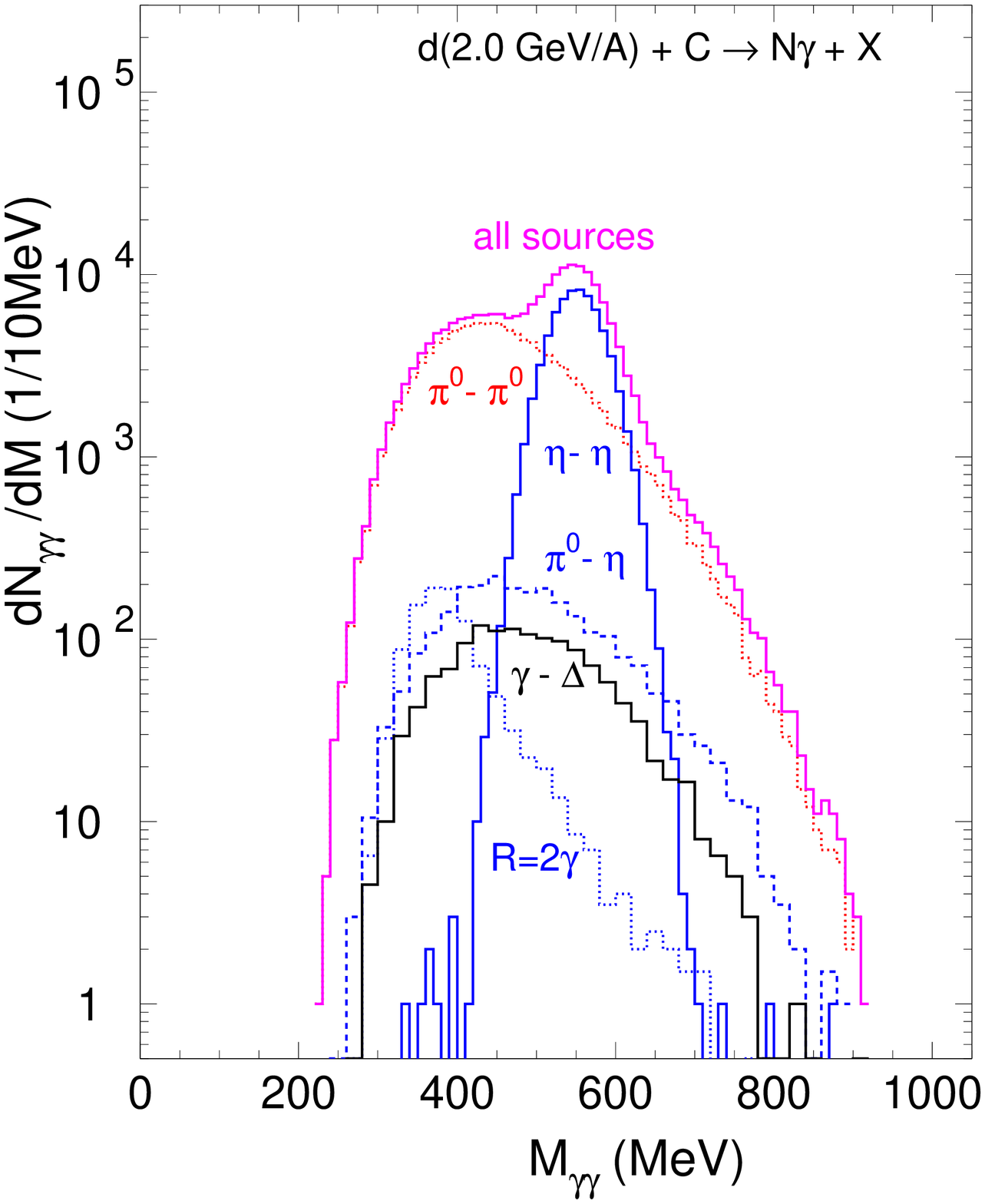}}
  \caption{(color online)
The calculated $\gamma\gamma$ invariant mass distribution  in $p$C
(left) and $d$C (right) collisions for selected  events with
$N_{\gamma}=2$. Contributions of different channels are shown.
Symbols near curves describe sources which photons  originate
from.
 }
 \label{MCcontr}
 \end{figure}

Model simulation allows us to disentangle different $\gamma$
sources to clarify the production mechanisms. This is illustrated
in Fig.\ref{MCcontr}. The double symbol near curves indicates the
sources where  both photons came from. The $\pi^0$ decay (marked
as $"\pi^0-\pi^0"$ in figures) naturally dominates in both the
reactions. In the energy range considered, the pion yield rapidly
increases and due to that a number of $\gamma\gamma$ pairs is
higher in the $p$C(4.6 GeV) than in the $d$C(2 AGeV) collisions.
The $\eta$ decay ($"\eta-\eta"$) is seen clearly, being spread
essentially due to uncertainties in the $\gamma$ energy
measurement. In the $d$C case the $\eta$  maximum is more
pronounced in the total distribution. It is of interest that the
$R$ resonance decay ($"R\to2\gamma"$) is also visible under
PHOTON-2 conditions. A number of $\gamma\gamma$ pairs from $R$ is
higher in the $p$C case, but the ratios for $\eta/R$ are
comparable: $\eta/R $= 34.18 and 24.95 for $d$C and $p$C
collisions, respectively. It means that a possibility to observe
the $R$ resonance depends on statistics of measured events.

One should note that since the low-mass enhancement in the
invariant $\pi\pi$ spectra showed  up clearly at beam energies
corresponding to the excitation of $\Delta$'s in the nuclear
system, the ABC effect was interpreted by a $\Delta\Delta$
excitation \cite{RS73,GFW99}. In particular, the early simplest
model for ABC production in $pn\to d+X^0$ involves the excitation
of both nucleons into $\Delta$-isobars through a one-pion exchange
where, after the decay of two $\Delta$'s , the final
neutron-proton pair sticks together to produce the observed
deuteron  \cite{RS73}. Though the enhancement observed in the
inclusive data for the $\pi^0\pi^0$ channel turns out in some
cases to be much larger than the predicted in these calculations,
the $\Delta\Delta$ mechanism is still attractive. More delicate
results on the vector and tensor analyzing powers in $\vec{d}d\to
^4$He+$X^0$ give strong quantitative support to this idea
\cite{Wea99}

The channel marked in Fig.\ref{MCcontr} as $"\gamma-\Delta"$
corresponds to the case when photons from the $\Delta$ decay
correlate with any other. Though the two-photon yield for this
channel in the $d$C case is close to that from the $R$ decay, the
maximum location is shifted to higher $M_{\gamma\gamma}$ by more
than 100 MeV. If one chooses both the photons from different
$\Delta$ isobars, for the PHOTON-2 conditions we have none event
from $10^9$  of simulated ones. One should note that we consider
incoherent $\Delta\Delta$ interactions and possible attraction in
this channel is not taken into account.

\subsection {The $\eta \to 3\pi^0$ decay}

As was indicated many years ago, a three-pion resonance having the
mass $M_\eta=$550 MeV and isospin $T=0$ maybe a pseudoscalar
particle of positive $G$ parity $(0^{-+})$. Under this assumption
it was shown that the partial rates and widths for this $\eta$
decay will be consistent with available experimental data
providing that the $\eta\to3\pi$ channel is enhanced by a strong
final state interaction \cite{BS62}. This strong interaction is
realized by postulating the existence of a particle having the
spin and parity $0^+$ and called a $\sigma$. Then the $3\pi$ decay
of the $\eta$ meson would proceed in two distinct steps:
$\eta\to\sigma+\pi^0$, $\sigma\to 2\pi^0$ or $(\pi^+\pi^-)$ where
the first step is an electromagnetic decay, while the second
occurs through the strong interaction. The fit to experimental
data gives the mass of a $\sigma$ particle about 370 MeV and a
full width of about 50 MeV \cite{BS62}. These parameters coincide
with those extracted from the direct analysis of pion spectra from
the $3\pi$ decay of the $\eta$ meson \cite{Crea63,Roy06}. The
enhancement of the $3\pi$ channel was argued by the presence of a
strong two-pion interaction which resembles the ABC effect. It is
of interest that the presence of such a pion-pion resonance
improves also the calculation of the $K_L-K_S$ mass difference
\cite{Ni64}.
\begin{figure}[h]
\centerline{\includegraphics[width=8cm]{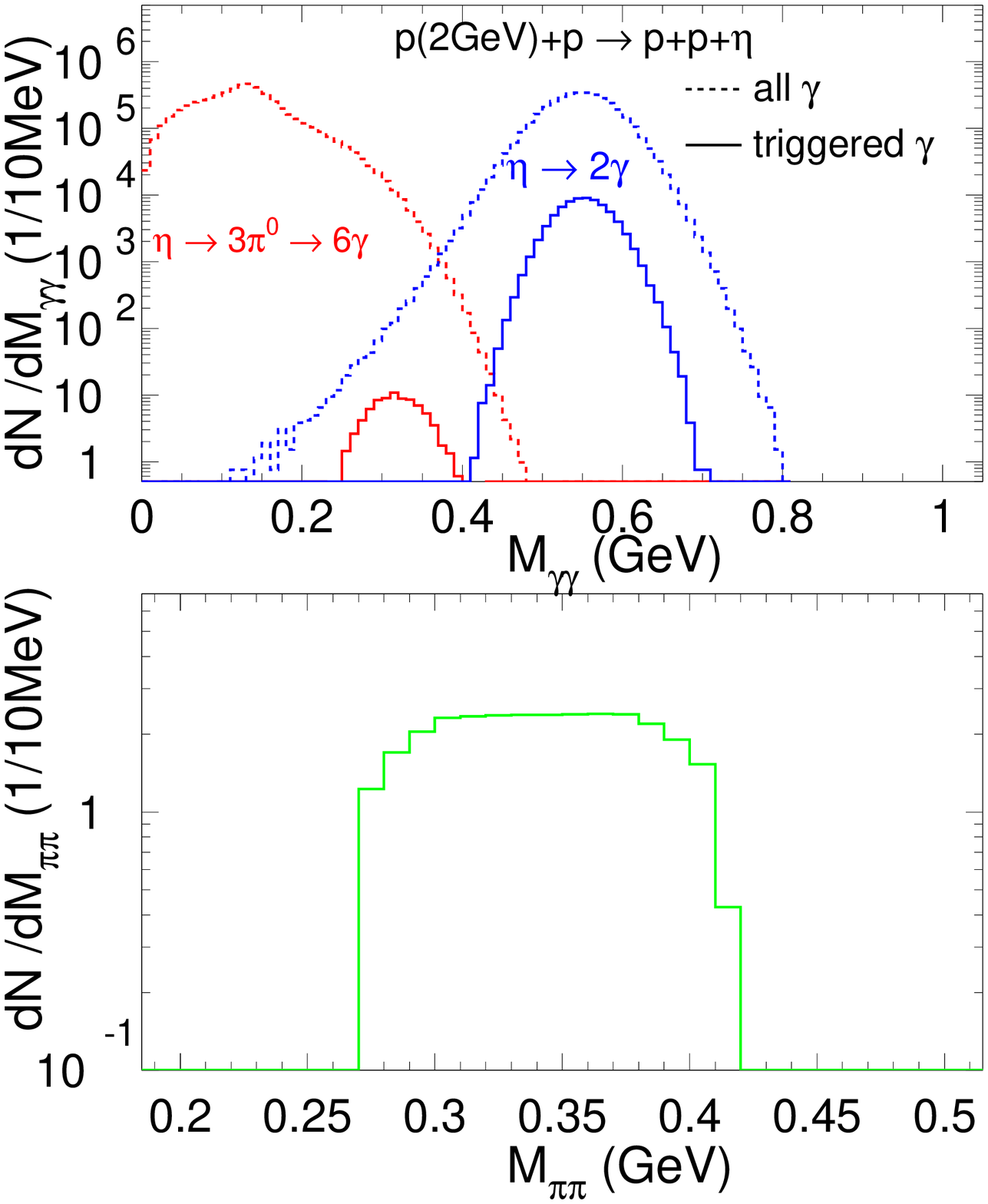}
\includegraphics[width=8cm]{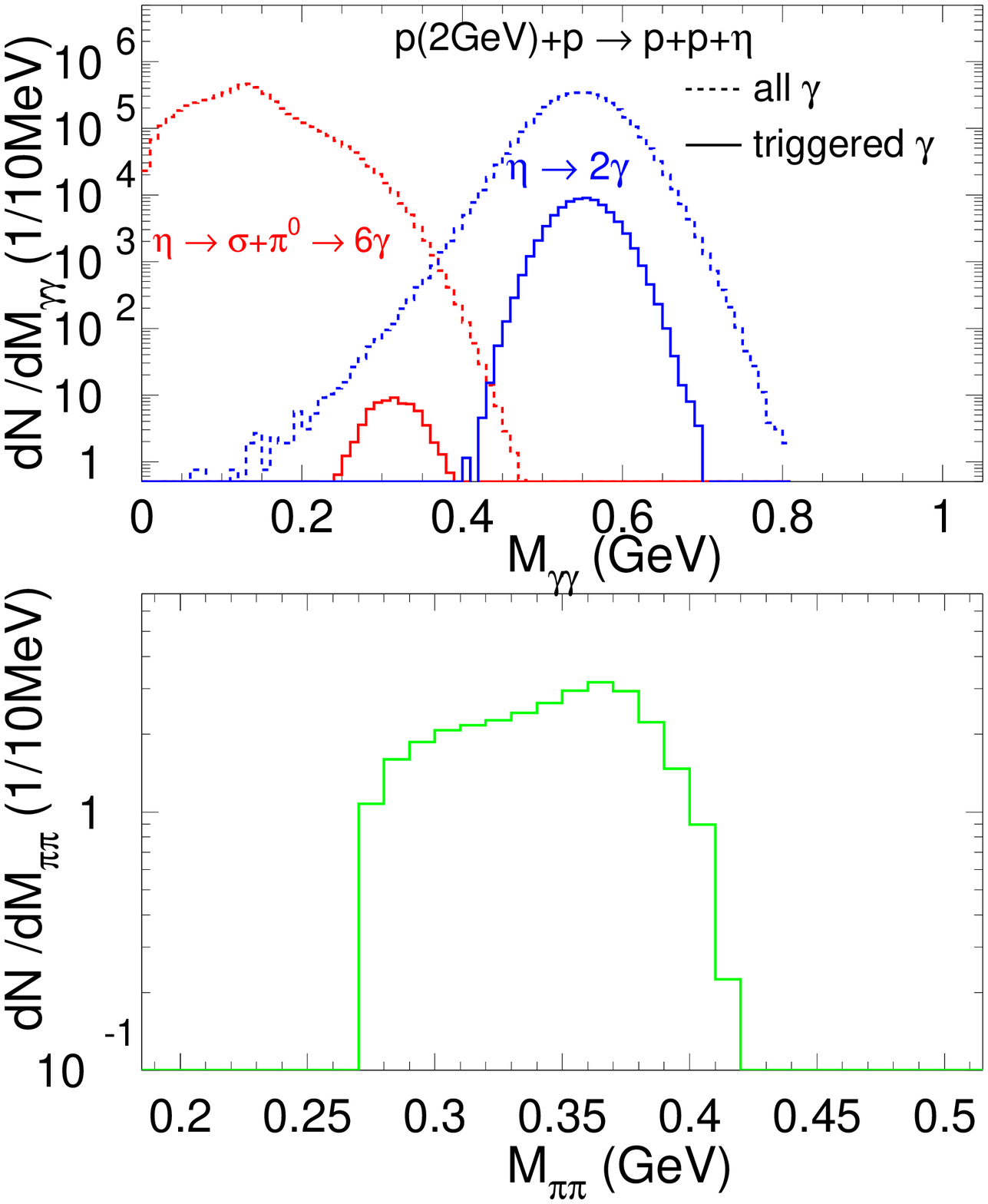}}
  \caption{(color online) The $\gamma\gamma$ (top) and $\pi\pi$
  (bottom) invariant mass
distributions for the $\eta$ decay through the $\eta\to
\gamma\gamma$ and $\eta\to 3\pi^0$ channels from pp collisions.
The left panels correspond to the $\eta\to 3\pi^0 \to 6\gamma$
mechanism, results in the right panels include interaction of two
pions via the $\sigma$ meson, $\eta \to \sigma +\pi^0 \to
6\gamma$. The dashed lines are calculated for the full $4\pi$
acceptance, the solid lines take into account the PHOTON-2
kinematical conditions. }
 \label{etadec}
\end{figure}

To check this mechanism (see Fig.\ref{etadec}) we simulated two
channels of the $\eta$ decay: the direct decay into two photons
$\eta\to 2\gamma$ and $\eta\to 3\pi^0$ which then decay into
photons. The last channel was calculated under two assumptions.
The first, all the three pions decay independently, $\pi^0\to
\gamma\gamma$, creating 6 photons. The second version assumes
that, as discussed above, two pions may interact forming $\sigma$
which decay into 2 pions, so $\eta \to \sigma +\pi^0 \to 6\gamma$.
As it is seen  (bottom panels in Fig.\ref{etadec}), the
interaction in the $\pi\pi$ channel results in some enhancement in
the $\pi\pi$ invariant mass spectra as compared the noninteracting
case. The $\gamma\gamma$ invariant mass spectra exhibit a spread
maximum near the pion mass and they are  practically identical in
both the cases (dashed lines). If the PHOTON-2 selection
conditions are implemented, a clear signal (solid lines in
Fig.\ref{etadec}) at $M_{\gamma\gamma}\sim 350$ MeV appears but
its intensity is by about three orders of magnitude lower than
that for the $\eta$ meson. It is of interest that the absolute
values and shape of the $M_{\gamma\gamma}$ spectrum are again very
similar in both the versions. So in the $\eta\to 3\pi^0$ decay it
is hardly ever possible to disentangle the cases with and without
the two-pion interaction by the detection of decay photons.

\subsection{Dibaryon mechanism}

Recently, a resonance-like structure has been found by the
CELSIUS-WASA Collaboration in the two-photon invariant mass
spectrum near $M_{\gamma\gamma} \sim 2m_\pi$ of the exclusive
reaction $pp\to pp\gamma\gamma$ at 1.2 and 1.36 GeV \cite{Bea05}.
This observation was interpreted as  $\sigma$ channel pion loops
which are generated by the $pp$ collision process and decay into
the $\gamma\gamma$ channel. Interference with the underlying
double bremsstrahlung background can give a reasonable account of
data\cite{Bea05}.

In Ref.\cite{KG07}, some arguments were given that such an
interpretation is at least questionable and an alternative
explanation was proposed where a possible origin of the structure
is based on the dibaryon mechanism of the two-photon emission
\cite{Kea01}. The proposed mechanism  $NN\to d_1^\star \to
NN\gamma\gamma$ proceeds trough a sequential emission of two
photons, one of which is caused by production of the decoupled
baryon resonance $d_1^\star$ and the other is its subsequent
decay. The $pp\to pp\gamma\gamma$ transition is treated in
\cite{KG07} within the assumption that at a large distance the
$NN$-decoupled six quark $d_1^\star$ state is a bound
$p\Delta(1232)$ state with the spin-parity  $J^P=0^-$ and isospin
$I=2$ \cite{M97}. The matrix elements were estimated
phenomenologically and effects of the final state interactions
between decay protons were included. This model reproduces
reasonably well the experimentally observed $M_{\gamma\gamma}$
spectrum of the $pp\to pp\gamma\gamma$ reaction in the vicinity of
the resonance structure \cite{KG07}.
\begin{figure}[h]
\centerline{\includegraphics[width=8cm]{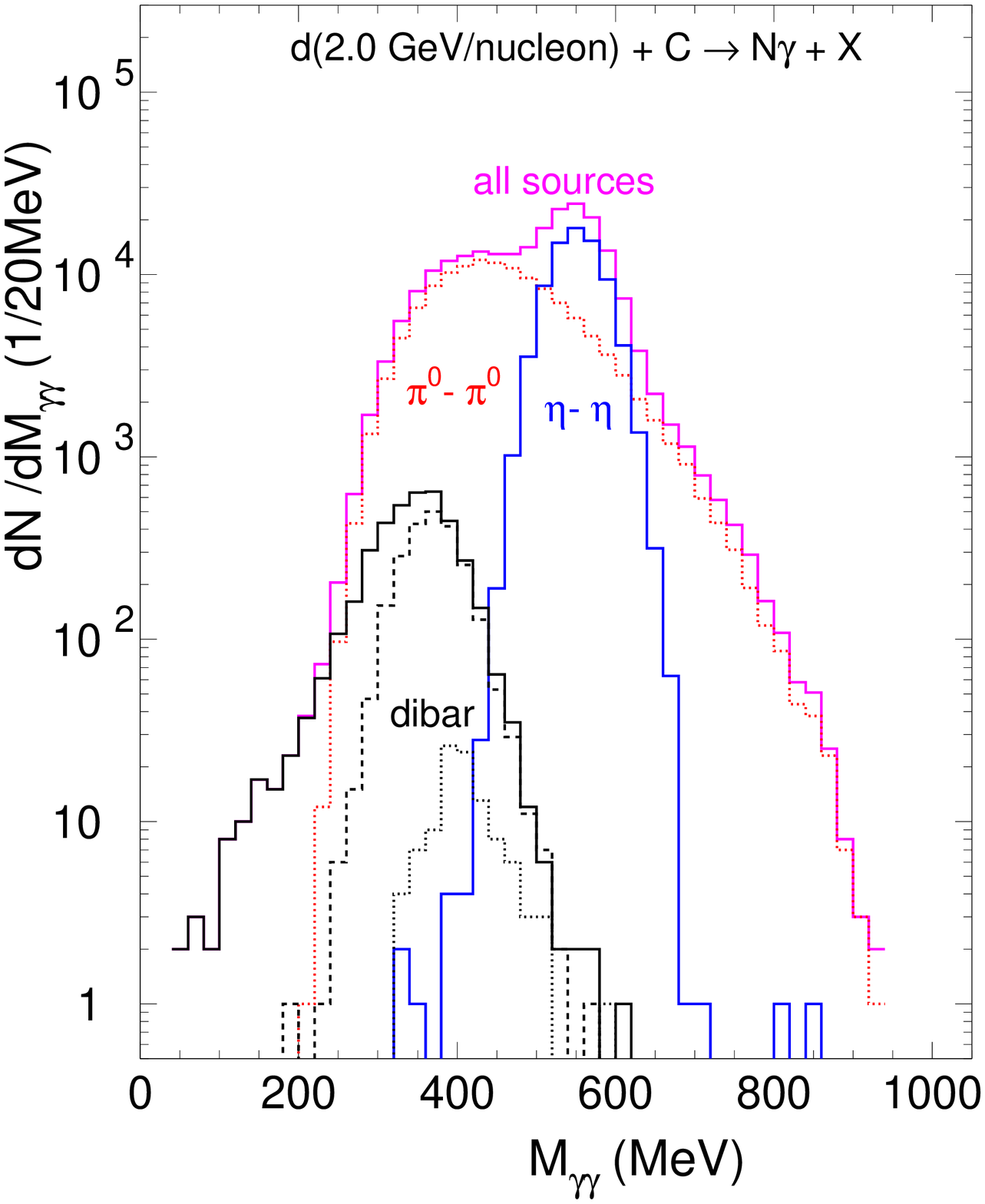}
\includegraphics[width=8cm]{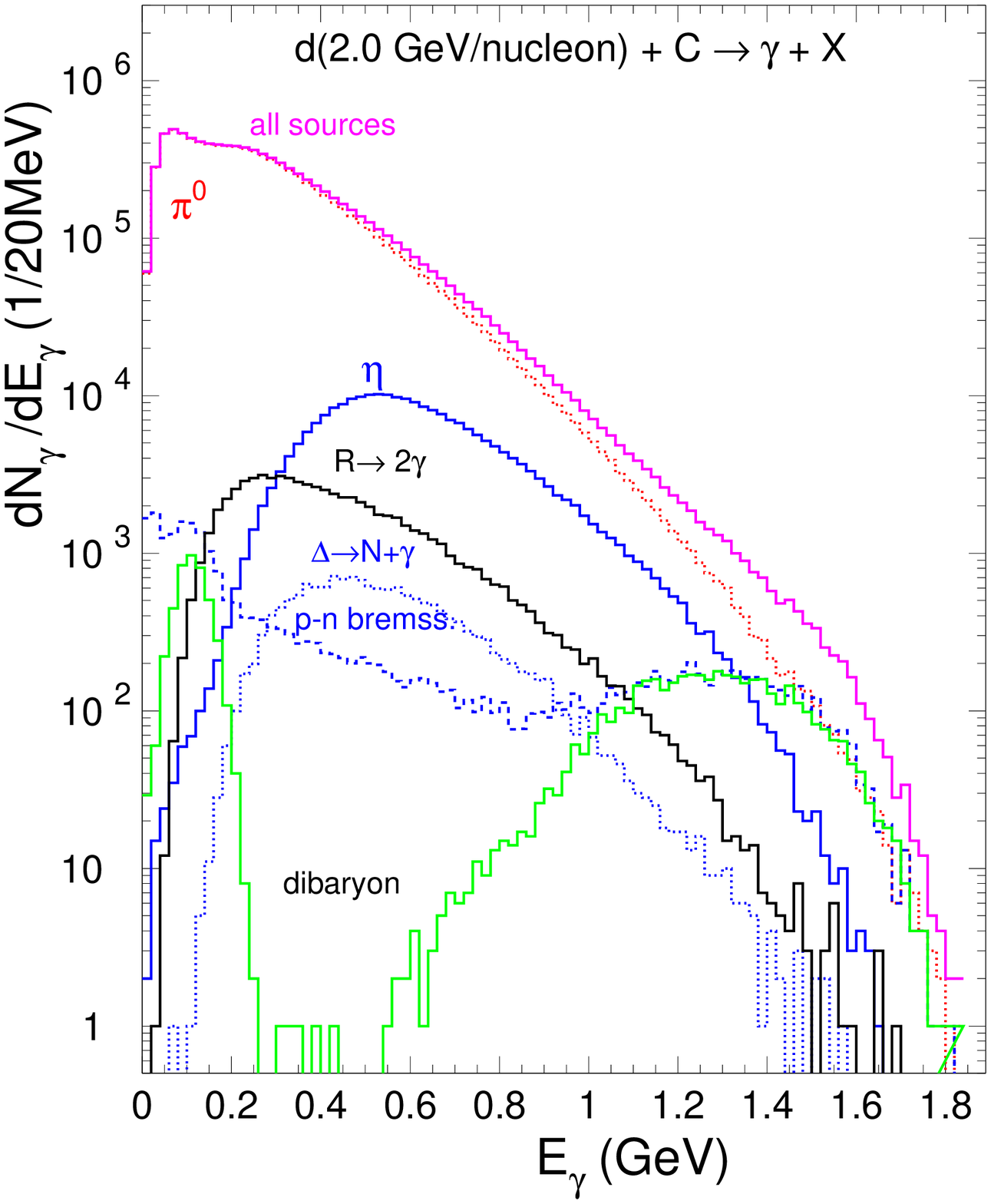}}
  \caption{(color online) The $\gamma\gamma$ invariant mass distribution (left)
and energy spectra of photons (right) calculated for $d$C
collisions with inclusion of the dibaryon mechanism. Contributions
of different channels are shown similarly to Fig.\ref{MCcontr}.
All curves beside the dibaryon one are given for the PHOTON-2
selected events. For the dibaryon channel (marked as "dibar") the
total two photon yield is presented.
 }
 \label{dCdibar}
\end{figure}

We would like to check whether this dibaryon mechanism may be
responsible for the peak observed in the $M_{\gamma\gamma}$
distribution of the $d$C collisions at $T=2$ AGeV. The two-step
scheme $NN\to d_1^\star \gamma \to NN\gamma\gamma$, where the
dibaryon mass is $m_d=m_N+m_\Delta =2.182$ GeV,  can be easily
simulated and included into our transport model. The only unknown
quantity is the cross section of this process. We estimated it by
means of the linear extrapolation of the two available points
measured at 1.2 and 1.36 GeV \cite{Bea05} till the energy of about
2 GeV.

In Fig.\ref{dCdibar}, such model calculation results with
inclusion of the dibaryon channel are presented. If the beam
energy of the $pp$ collision is fixed by 2 GeV, the photon energy
at the first step $pp\to d_1^\star \gamma$ will be a line, $
E_{\gamma 1}=(s-M_d^2)/(2\sqrt s )=640$ MeV  in the c.m. system.
Here $s$ is the total $pp$ colliding energy squared. In the lab.
system the maximal photon energy reaches about 1.5 GeV. However,
the photon from the second stage will be soft, being defined by
the baryon mass. Nevertheless, the maximum position of appropriate
two photon distribution moves also with the energy increase. The
total photon spectra from dibaryon and other sources from $d$C
interactions are shown in the right panel of Fig.\ref{dCdibar} and
it has a two-bump structure which mirrors the two-step production
mechanism. One should note that in contrast with all other results
in this figure the distributions of the dibaryon channel are
obtained for the full $4\pi$ acceptance without any limitation on
energy. If the PHOTON-2 selection conditions are implemented to
the dibaryon channel, the low energy photons are cut and we get no
two-photon pair from the dibaryon among $10^9$ simulated
collisions. Therefore, this dibaryon mechanism cannot explain the
observed anomaly.

One should note that in a certain sense the model considered is a
conventional dibaryon model where no nonhadronic degrees of
freedom are involved. Generally, these results may differ from
those obtained within nonconventional dibaryon models like
\cite{Kuk1}.

\section{Estimate of the $\eta$ and $R$ production cross sections
and resonance widths} The summed number of $p$C- and
$d$C-interactions in the experiment amounts to $\sim 3\cdot
10^{11}$ and $\sim 2\cdot 10^{12}$, respectively. The inelastic
cross sections of the observed $p$C and $d$C reactions  are
$\sigma_{\rm inel}(p{\rm C})=$411 mb  and $\sigma_{\rm inel}(d{\rm
C})=426$ mb \cite{Bar93}, respectively.

The cross section for the $\eta$ production in $d$C collisions
(similarly to $p$C interactions) is defined as follows:
\begin{equation}
\sigma(d{\rm C}\to \eta+X) = \sigma_{\rm inel}(d{\rm C}) \ \cdot \
\frac{N_\eta^{\rm exp} }{N_{d{\rm C-inter}}}  \cdot \ \frac{N^{\rm
mod}_{\rm all \ \eta}}{N_\eta^{\rm mod}/K_{\rm opt}}~.
 \label{eq3}
\end{equation}

Here the first two factors are the reaction cross section and the
measured mean multiplicity of $\eta$ mesons. A number of true $d$C
interactions resulting in the $\eta$ production is given as
\begin{eqnarray}
 N_{d{\rm C-inter}} = K_{\rm empty} \cdot K_{\rm beam-absorb}  \cdot
 N_d~,
 \label{eq4}
 \end{eqnarray}
where the total number of beam particles passing through the
target $N_d = (1 \div 2.5)\cdot 10^{12}$ is corrected on possible
interactions outside the target $K_{\rm empty}\sim 0.995$, to be
estimated by a special experiment with an empty target, as
discussed in Sect. \ref{sec2.2}, and the beam absorption $K_{\rm
beam-absorb}=0.5\pm 0.2$. The last factor reflects a particularity
of internal beam experiments where the direct monitoring is
impossible and part of the initial beam does not interact during
the working circle.

The third factor in eq.(\ref{eq3}) is the ratio of the total
number of simulated $\eta$ mesons to a number of $\eta$'s decaying
into two photons under PHOTON-2 experimental conditions. The
coefficient $K_{\rm opt}=8.93$ takes into account the rotation of
the modelled events in the $\phi$  plane to find other possible
$\gamma\gamma$ pairs in the given event satisfying the selection
conditions. Assuming that photons in the event are distributed
homogeneously in  $\phi$, this trick allows one to increase
effectively statistics of the selected events.

So for the $\eta$ production in $d$C collisions we have \\
  $$ \sigma(d{\rm C}\to \eta+X)
  = 1.31\pm 0.11^{+1.24}_{-0.89} \ {\rm mb}~. $$

In the case of  $p$C collisions we get $N_p = (1.5 \div 5)\cdot
10^{11}$, $K_{\rm beam-absorb}=0.8\pm 0.2$, coefficients
$K_{empty}$ and $K_{opt}$ are practically the same as for
$d$C-collisions,
so for the cross section we have\\
$$ \sigma(p{\rm C}\to \eta+X)=
 3.2\pm 0.2^{+4.5}_{-1.9} \ {\rm mb}~.$$
Large systematic errors of cross sections are coming mainly from a
problem of monitoring the intensity of the internal beam.

\begin{figure}[h]
\centerline{ \includegraphics[width=10cm]{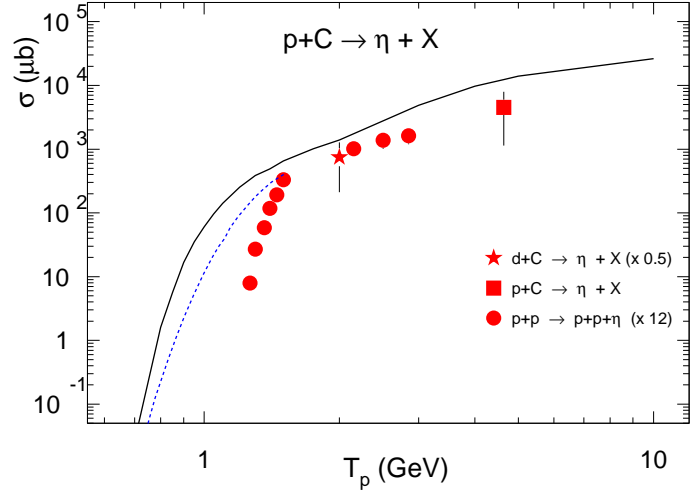}}
  \caption{(color online)Excitation function for $\eta$ production in $p$C collisions
calculated by Cassing~\cite{Cass} (dotted curve) and within our
DCM model (solid line).  Our measured points for $d$C (with a
factor of 0.5) and $p$C collisions are plotted by the star and
square, respectively. Circles show the cross section for
elementary collisions $pp\to pp \eta$ multiplied by the factor of
12.
 }
 \label{eta_sect}
\end{figure}

The obtained values for cross sections are compared in
Fig.\ref{eta_sect} with calculated excitation functions for the
$\eta$ meson production in the $p$C reaction. Note that the DCM
model agrees in the absolute scale with the experimental
differential cross section for this reaction in the energy range
of $T_p=0.8\div 2.0$ GeV (see Fig.\ref{pceta1}). The measured
$\sigma(p{\rm C}\to \eta+X)$ is below theoretical predictions by a
factor of about 2. The scaled $pp$ excitation essentially differs
from nuclear one in the near-threshold region due to Fermi motion.
The needed scale factor $A=12$ is a little bit higher than naively
expected A$^{2/3}$ because these data correspond to the $pp\to
pp\eta$ channel only while nuclear excitation functions include
all channels, $pp\to \eta+X$. Unfortunately, there is no other
experimental points in this energy range.

\begin{figure}[h]
\centerline{
\includegraphics[width=10cm]{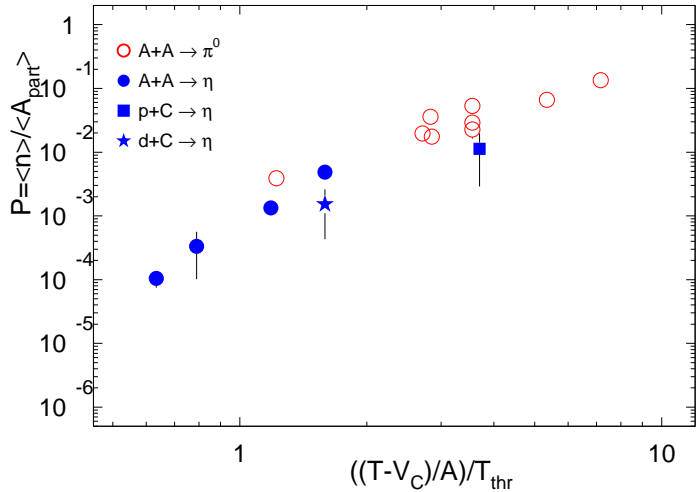}}
  \caption{(color online) Meson production probability per participant
  nucleon as a
function of the bombarding energy per nucleon  normalized to the
respective meson production threshold $T_{thr}$ in free
nucleon-nucleon collisions. Experimental points of the TAPS
collaboration are taken from the review \cite{K95}. Our points for
$d$C and $p$C collisions are plotted by the star and filled
square.
 }
 \label{comp}

\end{figure}

If the cross section for the $\eta$ production is known, the cross
section for the $R$-resonance may be estimated as follows:

\begin{eqnarray}
&&\sigma(d{\rm C}\to R\to\gamma\gamma)=\sigma(d{\rm C}\to \eta +X)
\ \cdot \ Br(\eta\to \gamma\gamma) \cdot \ \frac{N^{\rm exp}(R\to
\gamma\gamma)/\epsilon_R} {N^{\rm
exp}(\eta\to \gamma\gamma)/\epsilon_\eta}= \nonumber\\
&= &(0.075\pm 0.018)\cdot \sigma(d{\rm C}\to \eta+X)= 98\pm
24^{+93}_{-67}~{\rm \mu b}~,
 \label{eq36}
 \end{eqnarray}

where the branching ratio is $Br(\eta\to \gamma\gamma) =0.38$,
$\epsilon_R =N^{\rm mod}(R\to \gamma\gamma)/N^{\rm mod}_{tot}(R)$
and $\epsilon_\eta =N^{\rm mod}(\eta\to \gamma\gamma)/N^{\rm
mod}_{\rm tot}(\eta) $ are the  detection and selection
efficiency. The measured average reduced multiplicities are
compared in Fig.\ref{comp} with available systematics for meson
production near the threshold energies.
This simple energy scaling systematics for the subthreshold and
near threshold particle production was proposed in \cite{M93}. A
number of participant nucleons is estimated from the geometrical
consideration as
\begin{eqnarray}
<A_{part}>=\frac{A_p \ A_T^{2/3}+A_T \
A_p^{2/3}}{(A_p^{1/3}+A_T^{1/3})^2 }~.
 \label{Systematics}
\end{eqnarray}
The bombarding kinetic energy is corrected for the Coulomb barrier
$V_c$. This systematics is valid also for $K$ and $\rho$ mesons
\cite{K95,M93}. As is seen, the measured $\eta$ production points
are somewhat below the general trend. Partially, it may be caused
also by that all other experimental points correspond to heavier
nuclear systems and eq.(\ref{Systematics}) is not very justified
for our reactions.

As to the true internal width $w$ of the observed resonances, they
are defined by the measured width $w_{\rm measur}$  and also
specified by the spectrometer resolution $w_{\rm sp}$~:
 \begin{equation}
  w=(w_{\rm measur}^2 - w^2_{\rm sp})^{1/2}~.
  \label{w}
  \end{equation}
The spectrometer resolution depends on the $M_{\gamma\gamma}$
range. For the $R$ and $\eta$ invariant mass range we have,
respectively,
\begin{eqnarray}
2w_{\rm sp} (340<M_{\gamma\gamma}<360 {\rm MeV})&=&52.6~{\rm MeV}, \nonumber \\
2w_{\rm sp} (540<M_{\gamma\gamma}<560 {\rm MeV})&=&68.6~{\rm MeV}.
 \label{winstr}
\end{eqnarray}
So, according to eq.(\ref{w}), the intrinsic widths of detected
resonances are
\begin{eqnarray}
2w (\eta\to\gamma\gamma)&\approx& 0. \nonumber\\
2w(R\to\gamma\gamma)&\simeq& 63.7\pm 17.8~.
\end{eqnarray}

As is expected, the width of the $\eta$-meson practically equals
0, whereas it essentially differs from zero for the observed
resonance. The value of $2w$ in the Gaussian distribution
(\ref{eq2}) practically coincides with the width $\Gamma$ in the
Breit-Wigner function; thus, the intrinsic width of the observed
resonance structure is about $64\pm 18$ MeV.

\section{Concluding remarks}

Thus, based on a thorough analysis of experimental data measured
at the JINR Nuclotron and  statistics of $2339 \pm 340$ events of
$1.5\cdot 10^6$ triggered interactions of a record total number
$2\cdot 10^{12}$ of $d$C-interactions there was observed a
resonance-like enhancement at the mass $M_{\gamma\gamma}= 360\pm
7\pm 9$ MeV, with the width $\Gamma=64\pm 18$ MeV. The production
cross section $\sigma_{\gamma\gamma}\sim 98~\mu b$ is estimated
preliminary in the invariant mass spectrum of two photons produced
in $d$C-interactions at momentum of incident deuterons 2.75 GeV/c
per nucleon. A structure like this is not observed in the
$M_{\gamma\gamma}$ spectrum from $p$C (5.5 GeV/c) interactions
while the $\eta$ meson is clearly seen in both the cases. These
results, obtained by means of the mixing event background, are
confirmed by the wavelet analysis.

  Due to the use of  internal beams of the JINR Nuclotron,
totally more than $10^6$ triggered events at high discriminator
thresholds were recorded for every reaction during these
experiments. Smallness of the signal-to-background ratio in the
R-resonance mass range needs  very high statistics for
observation. This demand was not satisfied in previous experiments
and therefore no structure has been observed in this invariant
mass  range. As was noted above, the only most statistically
meaningful measurement of the invariant mass spectra was made by
the TAPS-collaboration but to resolve the resonance structure
discussed, a number of registered events  in the TAPS experiment
should be increased by an order of magnitude.

To certain extent this enhancement at $M_{\gamma\gamma} \sim
(2-3)m_\pi$ is similar to the puzzling ABC effect observed for
two-pion pairs from nucleon-nucleon and lightest nuclei collisions
at the near threshold energy. In the given work we see that it
exists in the $\gamma\gamma$ channel and measurements are extended
to a heavier system. It means that this resonance-like structure
is a quite stable object which even survives in the nuclear
surrounding.

To understand the origin of the observed structure, several
dynamic mechanisms were attempted: production of the hypothetic
$R$ resonance in $\pi\pi$ interactions during the evolution of the
nuclear collision, formation of the $R$ resonance with
participation of photons from the $\Delta$ decay, the $\pi^0\pi^0$
interaction effect in the $3\pi^0$ channel of the $\eta$ decay,
and a particular decoupled dibaryon mechanism. Unfortunately, none
of these mechanisms is able to explain the measured value of the
resonance-like enhancement, though they contribute to the
invariant mass in the region of interest.

The carbon target is really the heaviest one  used in experiments
where ABC-like structure has been observed. In contrast with all
other experiments considered here, one may expect some
manifestation of in-medium effects.  The prominent feature of the
$\eta$ meson is that the $\eta$-nucleon system couples dominantly
to the  $N^*(1535)(S^{11})$  resonance at the threshold energies.
Hence, due to the $\eta$ coupling to $N^*(1535)$-nucleon-hole
modes, one could expect the eta meson nuclear dynamics to be
sensitive to modification  of nucleon and $N^*$ properties in
medium. As was shown in \cite{JK08}, the $\eta$ spectral function
at normal nuclear density has a second maximum near $m_\eta \sim
400$ MeV which may be associated to a partial chiral symmetry
restoration. Its two-photon decay inside a nucleus might give a
rise to a maximum close to the measured value of $R$.
Unfortunately, we cannot perform transport calculations with
taking into account the in-medium modification of hadron
properties.

As was noted in Introduction, the recent data of the WASA-CELSIUS
Collaboration provide a strong support to the idea of a nontrivial
dibaryon state \cite{Sea06,Cea08}. An attractive candidate for its
realization may be a model of the intermediate $\sigma$-dressed
dibaryon \cite{Kuk1}. In this model the short-range $NN$
interaction, described by the standard $t$-channel $\sigma$
exchange between two nucleons, is replaced with the respective
$s$-channel $\sigma$ exchange associated with the intermediate
dibaryon production treated as a $\sigma$-dressed six-quark bag.
The strong scalar $\sigma$-field arises around the symmetric $6q$
bag, because the change in the symmetry of six-quark state in the
transition from the $NN$ channel to the intermediate dressed-bag
state. Due to a strong attraction of the $\sigma$ meson to quarks,
this intensive $\sigma$ field squeezes the bag and increases its
density.  The contribution of the $s$ channel mechanism would be
generally much larger than the conventional t-channel one due to a
resonance-like enhancement. The high quark density in the
symmetric $6q$ state enhances meson field fluctuations around the
multiquark bag and thereby partially restores the chiral symmetry.
Therefore, the mass of $\sigma$ meson gets much lower and has been
estimated to be the value $m_\sigma \sim 350\div 380$ MeV. In its
turn, it should enhance the near-threshold pion and double-pion
production \cite{Kuk1,Kuk2}. In addition, a large variety of
nuclear data, in particular properties of short- and
intermediate-range of $NN$ and $3N$ potentials, has been explained
within this model; however, still there is no direct quantitative
calculations of the ABC-like effects.

From the experimental side it is highly desirable to determine
more accurately the mass, width, and cross section of the observed
resonance structure by enlarging the acceptance. To verify the
above conclusions new experiments are required to be carried out
under conditions appropriate for detection of pairs of two photons
within the invariant mass interval of 300-400 MeV.
 In this respect experiments on proton and carbon targets
with proton and deuteron beams at the same energy per nucleon
would be very useful. Some scanning in the beam energy will
clarify the possible resonance structure of this effect. By
varying the opening angle of the PHOTON-2 spectrometer it is
possible to get information about momentum spectra of the produced
resonance-like structure which could be a delicate test of the $R$
production mechanism.

\vspace*{8mm}
 {\bf Acknowledgements}

We thank  S.B.~Gerasimov, E.E.~Kolomeitsev and E.A.~Strokovsky for
fruitful discussions and reading the manuscript. We are grateful
to A.S. Danagulyan, V.D. Kekelidze, A.S. Khrykin, V.I.~Kukulin,
V.A.~Nikitin, A.M.~Sirunyan, O.V.~Te\-ryaev, G.A.~Vartapetyan for
 discussions and valuable remarks. Furthermore, we would like to
 thank S.V. Afanasev,
V.V.~Arkhipov, A.S.~Artemov, A.F.~Elishev, A.D.~Kovalenko,
V.A.~Krasnov, A.G.~Litvinenko, A.I.~Mala\-khov, S.N.~Plyashkevich
and the staff of the Nuclotron for their help in conducting the
experiment, as well as B.V. Batyunya, A.V.~Belozerov and
A.G.~Fedunov for their help in analyzing data.

The work was supported in part by the Russian Foundation for Basic
Research, grant 08-02-01003-a and a special program of the
Ministry of Education and Science of the Russian Federation, grant
RNP.2.1.1.5409.\\[5mm]

{\bf{Appendix A. Continuous wavelets with vanishing moments } }
Here we would like to elucidate some details of the wavelet
analysis.

The family of continuous wavelets with  {\it vanishing moments}
(VMW) is presented here by Gaussian wavelets (GW) which are
generated by derivatives of the Gaussian function (\ref{oneD}).
For canonical Gaussian with $x_0=0; \sigma=1$ and $A=1$ one
obtains
\begin{equation}
G_n(a,b)\equiv G_n(x)=(-1)^{n+1}\frac{d^n}{dx^n}{e}^{-x^2/2},
 \label{Gaussw}
 \end{equation}
  where $n > 0$ is the order of the $g_n(x)$ wavelet.
The normalizing coefficients of these wavelets $C_{g_n}$ are $2\pi
(n-1)!$.

The most known in the GW family is the second order GW
$$G_2(x)=(1-x^2){\rm e}^{-\frac{x^2}{2}},$$ which is also
known as "the Mexican hat" \cite{wave}.

We use here also GW of higher orders, in particular,
\begin{eqnarray}
G_4(x)=(6x^2-x^4-3){\rm e}^{-\frac{x^2}{2}} \\
G_6(x)=\rm -(x^6-15x^4+45x^2-15){\rm e}^{-\frac{x^2}{2}}\\
 \label{G4-G8}
G_8(x)=\rm  -(x^8-28x^6+\rm 210x^4-420x^2+ 105){\rm
e}^{-\frac{x^2}{2}}
 \end{eqnarray}

It is a remarkable fact that the wavelet transformation of
Gaussian (\ref{oneD}) looks like the corresponding wavelet.
Therefore, the general expression for the n-th wavelet coefficient
has the following form (see derivations in \cite{gwave}):
 \begin{equation}
 W_{g_n}(a,b)g=\frac{A\sigma a^{n+1/2}}
{\sqrt{(n-1)!} \ s^{n+1}} \ g_n\left(\frac{b-x_0}{s}\right),
 \label{gausswavecoef}
 \end{equation}
 where we denote $s=\sqrt{a^2+\sigma^2}$.

The parameters $a$ and $b$ of continuous wavelets in
eq.(\ref{cwt}) are changing continuously, which leads to the
redundant representation of the data. In some cases, the above
mentioned GW properties are quite useful, in particular, this
redundant representation facilitates careful spectrum
manipulations. The price of this redundancy consists in slow speed
of calculations. Besides, all signals to be analyzed have in
practice a discrete structure.

It is noteworthy that GW should be used with some care since they
are nonorthogonal, which may disturb amplitudes of the filtered
signal after their inverse transform. In this respect the discrete
wavelet transform looks more preferable for many applications of
computing calculations with real data \cite{bib1} and deserves
special consideration. As was noted above, in our particular case
of the continuous WVM one can identify resonances without the
inverse transformation.

 Let us demonstrate this scheme by the $G_2(a,b)$ wavelet example.

According to  eq.(\ref{gausswavecoef}), one can obtain the maximum
(absolute) value of  $G_2$ for a Gaussian (\ref{oneD}) at the
shift point $b=x_0$ as
 \begin{equation}
 \max_b \ W_{G_{2}}(a,x_{0})g=\frac{A\sigma a^{5/2}}{(a^2+\sigma^2)^{3/2}}.
 \label{W2ax0}
 \end{equation}
In the wavelet domain of $G_2(a,x_0)$ this dependence looks  like
a simple curve with one maximum. To find it, one has to solve the
equation
$$\frac{d}{da} \max_b \ W_{G_{2}}(a,x_{0})g=0$$
The corresponding calculations give the position of maximum at the
scale axis as $a_{max} = \sqrt{5}\sigma$. Since the maximum
location in the $G_2$ domain is stable when the signal is
contaminated by some noise (see \cite{gwave}), one can use the
obtained point $x_0, a_{max}$ to start the fit. Although  the
maximum of a real contaminated signal is inevitably blurred over
some area in the wavelet space due to various distortions, it can
nevertheless be used as a good starting point for iterations
minimizing a fitting functional.

\end{document}